\documentclass[a4paper, onecolumn, superscriptaddress, aps, pra, 11pt, notitlepage, tightenlines, accepted=2026-01-12]{quantumarticle}
\pdfoutput=1
\usepackage{preamble}

\usepackage{tikz}

\renewcommand{\braket}[1]{\left\langle #1 \right\rangle}

\def\llangle{\langle\!\langle}
\def\rrangle{\rangle\!\rangle}

\def\sbra#1{\mathinner{\llangle{#1}|}}
\def\sket#1{\mathinner{|{#1}\rrangle}}
\def\sbraket#1{\mathinner{\llangle{#1}\rrangle}}
\def\sketbra#1{\mathinner{|{#1}\rrangle}\!\mathinner{\llangle{#1}|}}

\DeclareMathOperator*{\SumInt}{%
\mathchoice%
  {\ooalign{$\displaystyle\sum$\cr\hidewidth$\displaystyle\int$\hidewidth\cr}}
  {\ooalign{\raisebox{.14\height}{\scalebox{.7}{$\textstyle\sum$}}\cr\hidewidth$\textstyle\int$\hidewidth\cr}}
  {\ooalign{\raisebox{.2\height}{\scalebox{.6}{$\scriptstyle\sum$}}\cr$\scriptstyle\int$\cr}}
  {\ooalign{\raisebox{.2\height}{\scalebox{.6}{$\scriptstyle\sum$}}\cr$\scriptstyle\int$\cr}}
}
\usepackage[OT2,T1]{fontenc}
\DeclareSymbolFont{cyrletters}{OT2}{wncyr}{m}{n}
\DeclareMathSymbol{\Sha}{\mathalpha}{cyrletters}{"58}

\newtcolorbox{mybox}{colback=gray!10!white, colframe=gray}

\begin{document}
\title{Chasing shadows with Gottesman--Kitaev--Preskill codes}
\author{Jonathan Conrad}
\email{jonathan.conrad@epfl.ch}

\affiliation{Institute of Computer and Communication Sciences, {\'E}cole Polytechnique F{\'e}d{\'e}rale de Lausanne (EPFL), Lausanne CH-1015, Switzerland}
\affiliation{Dahlem Center for Complex Quantum Systems, Physics Department, Freie
Universit{\"a}t Berlin, Arnimallee 14, 14195 Berlin, Germany}

\author{Jens Eisert}
\affiliation{Dahlem Center for Complex Quantum Systems, Physics Department, Freie
Universit{\"a}t Berlin, Arnimallee 14, 14195 Berlin, Germany}
\affiliation{Helmholtz-Zentrum Berlin f{\"u}r Materialien und Energie, Hahn-Meitner-Platz 1, 14109
Berlin, Germany}

\author{Steven T.\ Flammia}
\affiliation{Department of Computer Science, Virginia Tech, Alexandria, USA}
\affiliation{Phasecraft Inc., Washington DC, USA}

\date{13/01/2026}

\maketitle
\begin{abstract}
We consider the task of performing shadow tomography of a logical subsystem defined via the Gottesman--Kitaev--Preskill (GKP) error correcting code. Our protocol does not require the input state to be a code state but is implemented by appropriate twirling of the measurement channel, such that the encoded logical tomographic information becomes encoded in the classical shadow. We showcase this protocol for measurements natural in continuous variable (CV) quantum computing. For heterodyne measurement, the protocol yields a probabilistic decomposition of any input state into Gaussian states that simulate the encoded logical information of the input relative to a fixed GKP code where we prove bounds on the Gaussian compressibility of states in this setting. 
For photon parity measurements, our protocol is equivalent to a Wigner sampling protocol for which we develop the appropriate sampling strategies.
Finally, by randomizing over the reference GKP code, we show how Wigner samples of any input state relative to a random GKP codes can be used to estimate any sufficiently bounded observable. 
\end{abstract}

\section{Introduction}
Recent years have seen steady progress in experimental realizations of quantum error correction.
On the one hand, experiments towards qubit-based quantum error correction have demonstrated impressive control of large quantum systems consisting of hundreds of physical qubits to encode and process encoded logical information~\cite{Bluvstein_2023}, as well the ability to encode quantum information into a single quantum harmonic oscillator beyond break-even via the \emph{Gottesman--Kitaev--Preskill} (GKP) code from the realm of bosonic quantum error correction~\cite{Sivak_2023, GKP, konno2023propagating}. 
On the other hand, quantum experiments are being conducted to understand practical capabilities of present noisy quantum devices, e.g., through variational quantum algorithms \cite{Cerezo_2021}, or to benchmark the readily accessible ``quantumness" \cite{arute2019quantum} through randomized sampling experiments. 
Aside from experimental progress, the design of error mitigation and smart post-processing techniques yields valuable insights into the design of future experiments and has specifically developed into a quest for learning properties of quantum states from randomly accessible snapshots. 

Any such effort makes sense only, however, 
if the anticipated state preparations or protocols are being implemented with high levels of accuracy. 
To ensure this, one usually has to resort to techniques of benchmarking, certification or
tomographic recovery \cite{BenchmarkingReview}. 
An interesting technique in this realm is the so-called classical shadow tomography protocol \cite{aaronson2018shadow, Huang_2020, Toolbox} that demonstrates just how little classical information needs to be extracted from any quantum state to reproduce the expectation values of a bounded number of suitably bounded observables. 
This family of protocols reconstructs a description of the input state as perceived by a random observable. The so obtained ``average-case description'' of the state becomes appropriate for many observables, so that the choice on what observables to focus on can be made later.

On the technical level, the (classical) shadow tomography protocol combines two ingredients: the fact that a channel-twirl of a 
\emph{positive operator valued measure} (POVM)
capturing a generalized measurement
projects it onto a channel with the fixed structure of a depolarizing channel and the existence of strong statistical anticoncentration bounds for medians of means estimation. 
The protocol proceeds by implementing a Clifford channel-twirl of a 
POVM that outputs samples over the reconstructed pointer states -- which are stabilizer states -- such that expectation values of observables over such samples match the expectation value of the depolarized input state and can be classically processed to yield the targeted expectation value. 
We will refer to such a group twirl involving a POVM a
POVM twirl. The classical post-processing necessary is informed by the structure of the depolarizing channel that resulted from the projection and is efficiently possible due to the stabilizer state structure of the samples. Medians of means estimation then yields a process to combine samples and prove bounds on the necessary sample complexity. 
It has been recognized in refs.~\cite{Chen_2021, Koh2022classicalshadows} that the projective nature of the channel twirl allows one to render the protocol robust to errors in the POVM, as any noisy version of the POVM would simply be projected onto depolarizing channel with amended parameters, which can be accounted for in post-processing. 
In fact, the projected POVM is simply one instance of a very well-structured informationally complete (generalized) POVM~\cite{Innocenti, Acharya, Ngyuen} can be used to implement the protocol. 

Bosonic quantum systems also offer a wide variety of POVMs beyond those exactly expressible in finite systems, such as heterodyne, homodyne, photon-counting, and photon parity measurements. 
Despite the richness of these POVMs, they can be similarly tamed into an effective channel with a simple structure by appropriate twirling.
Recent work \cite{iosue2022continuousvariable,  gandhari2022continuousvariable} has shown that twirling techniques are applicable for continuous variable quantum systems by only focusing only on certain energy-constrained subspaces.  
Prior work has treated this constraint as necessary due to the inaccessibility of random operations on this infinite Hilbert space.  

In this work, we develop shadow tomography protocols by focusing on logical subsystems prescribed by the GKP code. 
This also yields an effective finite subspace of the infinite dimensional CV Hilbert space and we show how effective shadow tomography protocols can be derived that reproduce logical expectation values of operators relative to the chosen GKP-codes. 
On the technical level, this is executed by twirling a CV-POVM over a set of random \textit{logical} Clifford gates, which has the effect that the logical action of the POVM becomes projected onto a depolarizing channel. 
This effect is revealed when the pointer states output by the protocol are evaluated in accordance to a decoder associated to the code. 
This structure reveals an interesting interplay between the physical structure of the system and its logical content. 
We identify different applications of the GKP-shadow tomography toolbox developed here by considering different choices of bosonic POVMs and finally show how a general shadow protocol for CV states can be obtained by combining our GKP-shadow tomography tools with a random choice of GKP codes. 

For example, when the protocol is executed using heterodyne measurement as POVM, the fact that GKP Clifford gates are represented by Gaussian unitary operations implies that the protocol outputs an ensemble of \textit{Gaussian states}, which contain the same logical information (relative to the chosen GKP code) as the given input state. 
Here the key upshot of our combination of techniques is that the statistical methods used in classical shadow tomography allow for the derivation of rigorous bounds on the number of such Gaussian states needed to faithfully retain the logical content of the input. 
While the bounds we derive scale exponentially in the system size, the key point of this result is that they are obtained \textit{without} knowledge of an analytical expression for the input state or strong assumptions of its physical structure. In the context of engineering GKP states for quantum computation this is particularly relevant, since many different analytical approximations to GKP states are used throughout the literature~\cite{MatsuuraEquivalence} and its specific physical structure finally depends on varying engineering details~\cite{Sivak_2023, konno2023propagating}.

\begin{figure}
    \centering
    \includegraphics[width=0.7\linewidth]{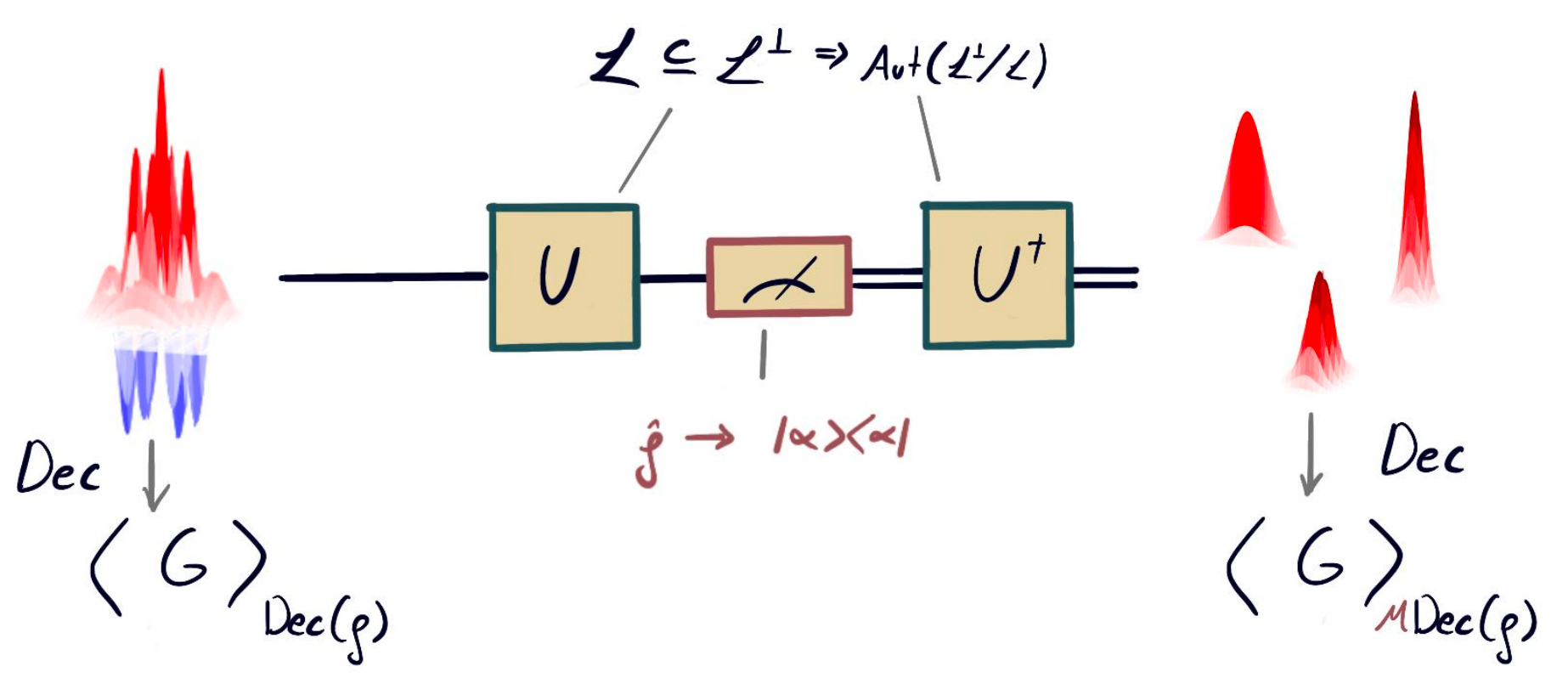}
    \caption{Illustration of Gaussian decomposition of arbitrary CV states via twirled heterodyne measurements. Relative to a GKP code described by a lattice $\CL\subseteq \CL^{\perp}$ is shadow tomography protocol yields a probabilistic decomposition of the input state into Gaussian states that reproduce logical expectation values up to a logical depolarization $\CM$.}
    \label{fig:heterodyne_shadow}
\end{figure}

This protocol yields an experimental \textit{black-box} procedure to convert an arbitrary physical input state into a convex combination of Gaussian states. 
As Gaussian states are easy to simulate classically \cite{Weedbrook_2012}, we expect this technique to be of value in assessing the performance of quantum computation and error correction using real GKP states. An illustration of this process is provided in fig.~\ref{fig:heterodyne_shadow}.
In the concrete application with photon parity detectors we consider, our protocol descends to one that samples the Wigner function of a given quantum state at random points according to a well-tailored distribution and we show how the toolbox we have developed allows one to derive sample complexity bounds for this protocol to estimate a number of arbitrary observables within certain bounds, unconditional on properties of the input state. 
This is achieved by combining the logical GKP shadow tomography protocol with a random choice of GKP code. Here the core technical ingredient is the existence of a Haar measure over the space of symplectic lattices and simple expressions for averages of functions of lattices over this measure. 
At the bottom, this is a randomized protocol to sample the Wigner function of an arbitrary CV state where the intricate way in which we choose where to sample allows us to rigorously bound the required sample complexity to estimate given CV observables to high confidence. 
By averaging over GKP codes, this protocol effectively interpolates between tomography of the \textit{logic} encoded in a physical system, and its \textit{physics}. 

The key point of this work, however, is to highlight the intersection between continuous variable physics and randomized tomographic methods originally derived for discrete variable systems. 
As becomes apparent in the course of our presentation, thinking about classical shadow tomography through the lens of GKP codes helps to refine our general understanding of the nuances of the classical shadow protocol while, vice versa, we obtain experimental handles to learn relevant aspects of a physical CV state using methods from random coding theory. 
Next to the presentation of the concrete results our purpose here is hence a pedagogical one: we hope that, through the lens of GKP codes, our explorations help the curious reader to develop a more refined understanding of the interesting intersection between physics, logic and everything in between.
\\
\subsection{Overview}
This article is structured as follows.  
We begin by a review of the basics of quantum harmonic oscillators and the structure of GKP codes in sec.\  \ref{sec:prelim}. 
In sec.\  \ref{sec:twirl}, we give a broad overview on twirling and discuss its various incarnations in state purification protocols, dynamical decoupling, noise mitigation and shadow tomography. 
This section is meant to provide a pedagogical ground-up introduction to the utility of twirling and the role of random operations for quantum experiments. 
We discuss extensions of these tools to the realm of continuous variable systems and explain how the infinitude of the associated groups of Gaussian unitaries can be suitably regularized. 
Finally, in sec.\ \ref{sec:GKPShadows} we apply the developed tools to design and prove bounds for logical shadow tomography protocols relative to GKP codes,  where we also examine the behavior of GKP shadow tomography with random GKP codes in sec.~\ref{sec:chasing_shadows}. 
We show this yields a protocol that allows one to estimate \textit{arbitrary} CV observables in a regularized manner. 
The core contribution of this section is the introduction of new techniques for estimating and bounding the performance of full tomography of a CV quantum system.
We close with a brief discussion and open questions for future work. 
For a better flow of presentation, the detailed proofs of most statements made throughout this manuscript are found in the appendix.

\section{Preliminaries}\label{sec:prelim}
In this section we review basic knowledge on continuous variable quantum physics and quantum error correction with the GKP code relevant to understand this manuscript. The following material is brief and introductory, and we refer the reader to the established literature, see, 
e.g., refs.~\cite{Weedbrook_2012, Gerry_Knight_2023, Terhal_2020} and references therein, for further details.

\subsection{Quantum harmonic oscillators}
Bosonic quantum error correction studies 
the robust embedding of discrete quantum information into a system of multiple \emph{quantum harmonic oscillators} (QHO), each of which can be described by an infinite dimensional Hilbert space $\CH=\operatorname{span}\lrc{\ket{n}}_{n=0}^{\infty}$ where $\ket{n}$ denote 
the well known Fock state vectors whose labels correspond to the \emph{eigenvalues} of the number operator $\hat{n}=\hat{a}^{\dagger}\hat{a}$ and $\hat{a}=(\hat{q}+i\hat{p})/\sqrt{2}$ denotes the annihilation operator.  $\hat{q}$ and $\hat{p}$ are the position and momentum operators, the canonical 
coordinates, whose improper eigenstates yield a basis for the underlying Hilbert 
space. The associated phase space inherits a non-trivial geometry from the canonical commutation relations (we will set $\hbar=1$ throughout),
and is most naturally studied in the Heisenberg 
frame,  i.e., in terms of the transformation behaviour of operators on this space. 
On a system of $n$ QHOs -- which we will refer to as having $n$ \textit{modes} --  we define a generalized quadrature operator  $\bs{\hat{x}}=\lr{\hat{q}_1, \hat{q}_2, \dots
, \hat{p}_{n-1} , \hat{p}_n }^T$ such that 
the canonical commutation relations are captured by the 
anti-symmetric symplectic form
\begin{equation}
J_n=\begin{pmatrix}
0 & I_n \\ -I_n & 0
\end{pmatrix}
\end{equation} 
where $I_n$ denotes the $n \times n$ identity matrix. Unless explicitly needed, we will omit the index $n$ from the symplectic form and simply denote it by $J$.

Analogous to the Pauli-operators for qubit-systems, the Heisenberg-Weyl operators for this infinite dimensional Hilbert space are given by displacement operators
\begin{align}
D\lr{\bs{\xi}} = \exp\left\{-i \sqrt{2\pi} \bs{\xi}^T J \bs{\hat{x}}\right\}\label{eq:displConvJ}
\end{align}
for $\bs{\xi} \in \R^{2n}$ being
elements of phase space. 
These displacement operators satisfy
the \emph{Weyl relations}
\begin{align}
D\lr{\bs{\xi}}
D\lr{\bs{\eta}}
= e^{-i\pi \bs{\xi}^TJ\bs{\eta}}
D\lr{\bs{\xi}+ \bs{\eta}}
\end{align}
for $\bs{\xi},\bs{\eta}\in \R^{2n}$
\footnote{The Weyl relations are actually a way of rigorously capturing the canonical commutation relations without having to resort to 
unbounded operators.}.
They form a basis for operators 
and are Hilbert-Schmidt orthogonal as $\Tr\lrq{D^{\dagger}\lr{\bs{\xi}}D\lr{\bs{\eta}}}=\delta^{(2n)}\lr{\bs{\xi}-\bs{\eta}}$, such that states can be represented by their Wigner function
\begin{equation}
W_{\rho}\lr{\bs{x}}
= \int_{\mathbb{R}^{2n}} d\bs{\eta}\, e^{-i2\pi \bs{x}^T J \bs{\eta}} \Tr\lrq{D\lr{\bs{\eta}} \rho}. \label{eq:Wigner}
\end{equation}

Displacement operators represent the unitary time evolution induced by Hamiltonians linear in the quadrature operators that implement the transformation \begin{equation}
D\lr{\bs{\xi}}^{\dagger} \bs{\hat{x}}D\lr{\bs{\xi}}=\bs{\hat{x}} + \sqrt{2\pi}\bs{\xi}
\end{equation}
 and commute and multiply 
 as
\begin{align}
D\lr{\bs{\xi}}D\lr{\bs{\eta}}&=
e^{-i\pi \bs{\xi}^TJ\bs{\eta}}
D\lr{\bs{\xi}+\bs{\eta}}, \\
&=
e^{-i2\pi \bs{\xi}^TJ\bs{\eta}}D\lr{\bs{\eta}}D\lr{\bs{\xi}}.
\nonumber
\end{align}
It is these properties that make them a natural set to choose stabilizer groups from.

Unitary evolution via Hamiltonians strictly quadratic in the quadrature operators, 
also termed \textit{Gaussian} unitary transformations \cite{Weedbrook_2012,Continuous}, implement symplectic transformations
\begin{align}
U=e^{-\frac{i}{2}\bs{\hat{x}}^T C \bs{\hat{x}}},\; C=C^T, \\
U^{\dagger}\bs{\hat{x}}U=S\bs{\hat{x}},\; S=e^{CJ}, \label{eq:sympH}
\end{align}
where $S\in \Sp_{2n}\lr{\R}=\lrc{S\in \R^{2n\times 2n} : S^TJS=J}$ is a real symplectic matrix which follows from unitarity of $U$ and we have 
\begin{equation}
D\lr{\bs{\xi}}U_S=U_SD\lr{S^{-1}\bs{\xi}},
\end{equation}
such that it also holds that
\begin{equation}
W_{U_S\rho U_S^{\dagger}}\lr{\bs{x}}=W_{\rho}\lr{S\bs{x}}.
\end{equation}

\subsection{GKP codes and their Cliffords}
\begin{figure}
\center
\includegraphics[width=.7\columnwidth]{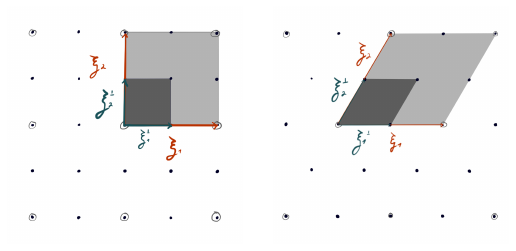}
\caption{The square $\Z^2$ (l.) and hexagonal $A_2$ (r.) GKP codes each encoding a qubit. The logical displacement amplitudes are marked in turquoise and stabilizer displacements are marked in red.}\label{fig:Z2_A2} 
\end{figure}

The GKP code \cite{GKP, Conrad_2022} -- or rather the family of it -- is a quantum error correcting code defined to embed discrete quantum information into a system of $n$ quantum harmonic oscillators by identifying a code space symmetric under the stabilizer group
\begin{equation}
\CS=\big\langle D\lr{\bs{\xi}_1}\hdots D\lr{\bs{\xi}_{2n}}   \big\rangle=\lrc{e^{i\phi_M\lr{\bs{\xi}}}D\lr{\bs{\xi}},\; \bs{\xi}\in \CL},  \label{eq:GKPdef}
\end{equation}
which is isomorphic to a full rank lattice $\CL= \mathbb{Z}^{2n}M$ with generator matrix
\begin{equation}
M=(\bs{\xi}_1,  \bs{\xi}_2,  \hdots ,  \bs{\xi}_{2n})^T.
\end{equation}
The symplectic dual lattice 
\begin{equation}
\CL^{\perp}=\lrc{\bs{x}\in \R^{2n}:\; \bs{x}^TJ\bs{\xi} \in \Z\, \forall \bs{\xi} \in \CL}
\end{equation}
labels the centralizer of the GKP stabilizer group,  such that the GKP stabilizer group is abelian if and only if it is isomorphic to a \textit{weakly symplectically self-dual} lattice
\begin{equation}
\CL \subseteq \CL^{\perp} \Leftrightarrow M=AM^{\perp},
\end{equation}
where the right hand side describes the sublattice structure by identifying how basis vectors of $\CL$ are described by (integer) linear combinations of basis vectors of $\CL^{\perp}$ as given by the
the symplectic Gram matrix $A=MJM^T$ when the dual basis is chosen via some canonical choice \cite{GKP}.  
The phases $\Phi_M\lr{\bs{\xi}}=\pi \bs{a}^T A_{\lowertriangle}\bs{a},\, \bs{a}=M^{-T}\bs{\xi} $ in eq.\ \eqref{eq:GKPdef} are determined by the basis in which the stabilizer generators are fixed to a $+1$ eigenvalue and are trivial when the symplectic Gram matrix $A$ is even \cite{Conrad_2022}.

A special class of GKP codes, called \textit{scaled} GKP codes, is obtained from rescaling a symplectic self dual lattice $\CL_0=\CL^{\perp}_0 \mapsto \CL=\sqrt{d}\CL_0:\, \CL \subseteq\CL^{\perp}, \;  $ via the square root of the desired local dimension $d\in \N$ and gives rise to the well-known GKP codes that encode a qubit ($d=2$) into a single oscillator via the square- or the hexagonal lattice with bases

\begin{equation}
M_{\Z^2}=\begin{pmatrix}
1 & 0 \\ 0 & 1
\end{pmatrix},\,
M_{A_2}=\frac{1}{\sqrt{2\sqrt{3}}}\begin{pmatrix}
2 & 0 \\ 1 & \sqrt{3}
\end{pmatrix}.
\end{equation}

These GKP codes have been widely explored in the literature: They afford a distance (given by the length of the shortest non-trivial logical
displacement) $\Delta\lr{\sqrt{2}\Z^2}=2^{-\frac{1}{2}}$ and $\Delta\lr{\sqrt{2}A_2}=3^{-\frac{1}{4}}$. We depict their structure in fig. \ref{fig:Z2_A2}, where it can also be seen that the lattices are respectively symmetric under $\pi/2$ and $\pi/3$ rotations $R_{\pi/2}$ and $R_{\pi/3}$ which correspond to the logical Hadamard $\hat{H}$ gate for the square GKP code and the Hadamard-phase gate $\hat{H}\hat{S}^{\dagger}$ for the hexagonal GKP code. 

The identification as logical Clifford gates is made through their property
as symplectic lattice automorphisms, of which the general structure has been explored in refs.~\cite{garden, burchards2024fiberbundlefaulttolerance, Royer_2022, GCB}. Another such symplectic
automorphism is given by the transvection $S=I+\bs{e}_{1}\bs{e}_2^T$, where $\bs{e}_i$ are unit vectors, that yields a logical phase gate $\hat{S}$ for the square GKP code. 
In general, for scaled GKP codes all symplectic autmorphisms are given by the symplectic matrices $S=M^TU^TM^{-T}$, where $M$ is the generator for the lattice basis and $U\in \Sp_{2n}\lr{\Z_d}$ labels the logical action of the corresponding non-trivial Clifford gate. We refer to ref.~\cite{garden} for an in-depth discussion. In the appendix, we show that any such $U\in \Sp_{2n}\lr{\Z_d}$  (and consequently, by conjugation with $M^T$, any corresponding real symplectic automorphism $S$) can be generated by a length $O(dn^2)$ sequence of elementary local matrices in $\Sp_{2n}\lr{\Z_d}$ that correspond to the qudit versions of the usual Hadamard, phase  and CNOT gates. Throughout this manuscript, by (logical) Clifford group we mean the group of automorphisms of logical Pauli operators under conjugation while the prefix \textit{trivial} refers to those elements that implement a projectively trivial automorphism. We refer to ref.~\cite{garden, burchards2024fiberbundlefaulttolerance} for a more in-depth discussion of these groups and their connection to the respective groups of lattice automorphisms.

\section{Twirling theory}\label{sec:twirl}

We now turn to discussing constructions and applications of \textit{random} logical GKP Clifford gates.  While random Clifford gates is a widely and well studied topic for qubit-based systems,  for the GKP code the question of how to define a measure over the -- now infinite -- trivial-  and non-trivial group becomes more nuanced.  
In this context, as a versatile tool, twirls become important as averages over groups. They here appear both as twirls on the logical- and on the physical level and will be the key technical ingredient in the following developments.

\begin{figure*}
\center
\includegraphics[width=.75\linewidth]{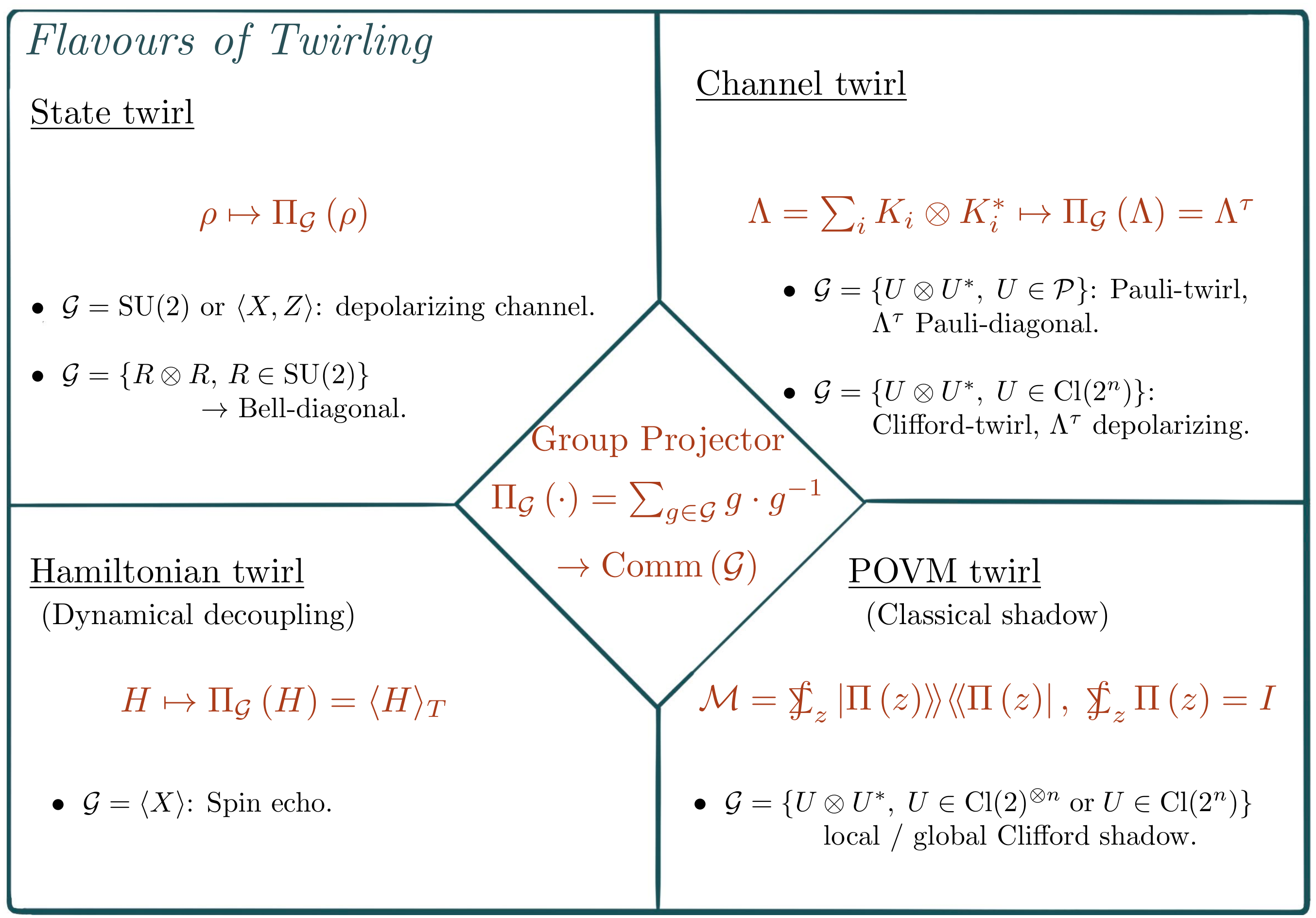}
\caption{Different notions of twirling discussed in the main text. }\label{fig:twirl}
\end{figure*}

Constructions of random (trivial or non-trivial) Clifford gates find widespread applications from state preparation to error mitigation to benchmarking and most notably in many recent works on shadow tomography.  The common ground of these applications is that the random Clifford gates are used to implement various incarnations of group projectors -- also phrased \textit{twirling} -- which we summarize in fig.~\ref{fig:twirl}. 
As exploited in ref.~\cite{Conrad_2021} for the square GKP code to Floquet-engineer a GKP Hamiltonian, once a suitable measure for one type of twirl is found, it is easy to translate it into the different incarnations. In the following we discuss the twirling theory for states, channels and POVMs. Although listed in fig.~\ref{fig:twirl}, we do not discuss the Hamiltonian twirl but merely mention its equivalence to the state twirl and refer to the relevant literature, see refs.~\cite{Zanardi_1999, Conrad_2021}.

\subsection{Twirling states, channels and POVMs} 

In this section, we first review how the different incarnations of twirling work on a qubit level before discussing its logical analogue using the GKP code. 

\paragraph{State twirling.} 
On the (logical) qubit level,  a state twirl over a group $\CG \subseteq \CU\lr{N}$ maps
\begin{equation}
\rho \mapsto \Pi_{\CG}\lr{\rho}=\int_{\CG} d\mu^{\CG}\lr{U}\,  U\rho U^{\dagger},\label{eq:group_proj}
\end{equation}
where the integration is taken over the group
of all unitaries and we specify the uniform measure $\mu^{\CG}\lr{U}$ over unitaries in $\CG$ (and zero outside). It can be verified that the corresponding group projector indeed is a projector $\Pi_{\CG}^2=\Pi_{\CG}$ and that it maps onto the \textit{commutant} $\lrc{U\in \CU\lr{N}:\; [U,g]=0\,  \forall g \in \CG}$ which is spanned by the \emph{irreducible representations} (irreps) of $\CG$.  Note that in eq.\  \eqref{eq:group_proj} phases attached to each group element $g$ do not matter and we in fact only need that the operators used in \eqref{eq:group_proj} form a projective unitary representation of the group.  The group projector associated to the Pauli group $\CP=\langle \hat{X},  \hat{Z} \rangle$ can hence be written as
\begin{equation}
\Pi_{\CP}=\Pi_{\langle \hat{X}\rangle} \circ \Pi_{\langle \hat{Z}\rangle}
\end{equation}
and projects any input state onto a state that is invariant under both random bit- and phase-flips.  The only state satisfying this property is completely mixed $\rho=I/2$,  aligned with the fact that the Pauli group forms a unitary $1$-design -- i.e.,  group twirls involving single powers $g,  g^{\dagger}$ as in eq.\ \eqref{eq:group_proj} act on a state in the same way as a twirl over the whole unitary group $\CG=\CU\lr{N}$ which has only the identity $I$ in its commutant. 

The same idea also works for subgroups, giving rise to twirls that project onto suitable symmetric subspaces. 
For example, we can consider state twirls for a group $\CG \subseteq \CU\lr{N}$ that acts as
\begin{equation}
\rho \mapsto \Pi_{\CG}\lr{\rho}=\int_{\CG} d\mu^{\CG}\lr{U}\,  (U\otimes U) \rho (U \otimes U)^{\dagger},
\end{equation}
where the integration is taken over the Haar measure over $\CG$. This is a group projector onto the commutant $\lrc{O\in L\lr{\CH\otimes \CH}:[O,g\otimes g]=0\,  \forall g \in \CG}$,
where $L\lr{\CH\otimes \CH}$ denotes the space of all linear operators on $\CH\otimes \CH$.
This form of the twirl has been applied in protocols to distill entangled states , where the random application of correlated Pauli operators ($\CG=\CP$) yields entangled (Werner) states \cite{Werner89,Bennett_1996}.

\paragraph{Channel twirling.} 

In a similar way as one can think of state twirls, it makes sense to consider channel twirls.
A channel twirl $\tau_{\CG}$ maps a quantum channel,  written in its $\chi$-matrix representation with $P_{\alpha}$ a Pauli operator with index $\alpha$
\begin{equation}
\CC\lr{\cdot}=\sum_{\alpha, \beta}\chi_{\alpha, \beta}P_{\alpha}\cdot P_{\beta} \label{eq:Lambda_chi}
\end{equation}
onto a ``symmetrized'' version of itself.   Specifically,  it is defined 
as the map
\begin{align}
\tau_{\CG} \circ \CC=\int_{\CU\lr{N}} d\mu\lr{U}\,   U\CC\lr{U^{\dagger}\cdot U}U^{\dagger}.
\end{align}
In vectorized notation where $\widehat{\CC}$ acts on state vectors $\sket{\rho}\mapsto \widehat{\CC}\sket{\rho}$,  this is recast in the recognizable form 
\begin{align}
\tau_{\CG} \circ \widehat{\CC}=\int_{\CU\lr{N}} d\mu\lr{U}\,  \CU \widehat{\CC} \CU^{\dagger}, \;  \CU:=U\otimes U^*
\end{align}
and again for the uniform measure $\mu^{\CG}$ on $\CG$ we can recognize that the channel is projected onto the commutant of the group representation $R\lr{\CG\otimes \CG^*}= \lrc{R\lr{g}\otimes R(g)^*\,  \forall g \in \CG}$.  

A Clifford twirl on a channel can be understood as a combination of twirls  
\begin{equation}
\Pi_{\Cl\lr{N}}=\Pi_{\Cl\lr{N}/P\lr{N}} \circ \Pi_{P\lr{N}}
\end{equation}
over the trivial- and non-trivial Clifford groups \cite{Dankert_2009,  DiVincenzo_2002},
where the channel twirl enforces that every pair of Pauli operators $P_{\alpha} \otimes P_{\beta}^*$ commute with every operator of the form $P\otimes P^*,  P\in \CP\lr{N}$.  Since Pauli operators only commute up to phases,  this twirl effectively enforces that the channel only has ``diagonal'' Pauli elements for which $\alpha=\beta$ in eq.\  \eqref{eq:Lambda_chi}.
Subsequently,  the twirl over the non-trivial Clifford elements enforces permutation symmetry on the non-trivial Pauli elements indexed in $\chi_{\alpha, \alpha}, \alpha \neq 0$.  For a single qubit,  the desired measure over non-trivial Cliffords hence should satisfy (up to phases)
 \begin{equation}
 \mu_{\Cl\lr{2}/\mathcal{P}} ( I\neq P \in \CP \mapsto 
X,  Y,  Z) =\frac{1}{3}, 
\label{eq:Clifftwirl}
 \end{equation}
and by assuming the input channel to be trace preserving,  the twirl produces a channel
\begin{equation}
\Pi_{\Cl\lr{N}}\circ \CC \lr{\rho}=\chi_{0,0}\rho+\frac{1-\chi_{0,0}}{3}\sum_{\alpha>0}P_{\alpha} \rho P_{\alpha}.
\end{equation}
Writing
\begin{equation}
\sum_{\alpha>0}P_{\alpha} \rho P_{\alpha}=\sum_{\alpha}P_{\alpha} \rho P_{\alpha} - \rho = 4 \Pi_{\CP}\lr{\rho} - \rho, 
\end{equation}
we obtain, knowing that the Pauli state-twirl produces a completely mixed state $\Pi_{\CP}\lr{\rho}=2^{-1} I$,
\begin{equation}
\Pi_{\Cl\lr{N}}\circ \CC \lr{\rho}=\frac{4 \chi_{0,0}-1}{3} \rho + \frac{2(1-\chi_{0,0})}{3}I.\label{eq:CL_twirled_channel}
\end{equation}
A channel twirl over the Clifford group $\Pi_{\Cl\lr{2}}$ that produces an output of such a form  -- i.e., it is in vectorized notation a linear combination of the trivial channel $I\otimes I^*$ and the fully depolarizing channel $\sketbra{I}$ -- is also called a Clifford $2$-design
\cite{Gross_2007}. 
There are several ways to implement a twirl over either the full or non-trivial Clifford group. 
One straightforward way to define a non-trivial Clifford twirl is to pick one of the $6$ elements in $\Sp_2\lr{2}$ at random  (corresponding to the matrices with $1$ entries on the diagonal and anti-diagonal and one of the four choices of $\Z_2^{2\times 2}$ matrices with one $0$ entry).  
While this strategy is the simplest,  it requires to exhaustively enumerate all elements of the non-trivial Clifford group does not generalize easily to a multi-qubit setting where it becomes desirable to approximate such twirl via a random walk over $\Sp_{2n}\lr{\Z}$ or,  more commonly,  to directly sample random Clifford circuits (involving Pauli gates) gate-by-gate to set up a good approximation to a (Haar-)random Clifford circuit.
On a single qubit,  the first strategy effectively samples from one of the $24$ elements in the set 
 \begin{align}
 S_0 =\{CP,\; C\in \Cl\lr{2}/\CP, \; P\in \CP \} \subseteq \C^{2\times 2}
 \end{align}
which, in fact, 
turns out to be larger than necessary. 
From the discussion above we have learned that the main function of the non-trivial Pauli twirl is to set up a random permutation of the Pauli operators. 
This function is already fulfilled by the $3$ element cyclic subgroup $\langle \hat{H}\hat{S}^{\dagger}\rangle$, so that only a total of $12$ elements are necessary to consider to build a single qubit Clifford twirl.  
In fact, it is easy to verify that the set
\begin{equation}
 S=\langle \hat{H}\hat{S}^{\dagger}\rangle \CP=\{ CP  \vert \; C\in\langle \hat{H}\hat{S}^{\dagger}\rangle, \; P\in \CP \} 
\end{equation}
is a Clifford 2-design. 
This is verified either by checking that it correctly twirls a Pauli-diagonal channel into a depolarizing channel from its transitivity over the Pauli operators or by verifying an equivalent condition proven in ref.~\cite{Gross_2007}, which is that the 
\textit{frame potential} evaluates to a value of $\CF=2$, proving that one indeed encounters an exact unitary 2-design.
In fact, the group stated above has the minimal cardinality of $(d^2-1)d^2$ shown to yield a Clifford 2-design \cite{Gross_2007, Steve_comment}, which, in our construction, comes from the fact that $\langle \hat{H}\hat{S}^{\dagger}\rangle$ is the minimal Pauli transitive subgroup. 
For any dimension, a Pauli transitive subgroup has at least $d^2-1$ elements, such that a Clifford design of the type proposed above has at least $(d^2-1)d^2$ elements, matching the bound conjectured ref.~\cite{Gross_2007} and later proven to hold in ref.~\cite{Steve_comment}. 

\paragraph{POVM twirling.} 

A special application of the channel twirl -- and at the same time a most important one for the purposes of this work --
is found when it is applied to a POVM reflecting a generalized measurement.  
This setting has been recently popularized by showing how it can be used for 
\textit{shadow tomography} \cite{aaronson2018shadow,  Huang_2020, Chen_2021,RBRobustShadows} giving rise to various \emph{noisy-intermediate-scale-quantum} (NISQ) friendly applications.  
On a single mode consider the POVM representing a 
computational basis measurement
\begin{equation}
\CM_Z=\sum_{z\in \Z_2} \sketbra{\Pi_z},\;  \Pi_z=\ketbra{z}.
\end{equation}
The corresponding $\chi$ matrix representation has $\chi_{0,0}=1/2$,  such that we can compute
\begin{equation}
\CM := \Pi_{\Cl\lr{2}}\circ \CM_Z = \lr{\rho + I} /3. 
\end{equation}
The twirled POVM is a depolarizing channel and is (as a matrix,  
not physically as a quantum channel) 
invertible with $\CM^{-1}\lr{X}=3X-I$. Furthermore,  
its action on state vectors $\sket{\rho}$
 \begin{equation}
 \CM\sket{\rho}=\frac{1}{|\Cl\lr{2}|}\sum_{C\in \Cl\lr{2} \\ z \in \Z_2} \CC \sketbra{\Pi_z} \CC^{\dagger} \sket{\rho}
 \end{equation}
can be interpreted as protocol to decompose 
arbitrary quantum state vectors $\sket{\rho}$ into stabilizer state vectors $\CC \sket{\Pi_z}$. Note that, as we have done earlier as well, we use the calligraphic symbols for unitary channels in place of $\CC=C \otimes C^*$.  An experimental protocol to reconstruct (properties of) the state is hence identified by realizing that measuring in the computational basis after applying random Clifford gates to an input state allows for a reconstruction
\begin{equation}
\sket{\rho}=\CM^{-1}  \mathbb{E} \lrq{ \CC_i \sket{\Pi_{z_i}}}, \label{eq:approx_state}
\end{equation}
where the Cliffords $C_i$ are picked uniformly from the Clifford group and $z$ is determined by the Born rule $ z_i \sim  \sbraket{\Pi_z | \CC^{\dagger} | \rho}$.  Similarly,  expectation values of observables can also be estimated as
\begin{equation}
\sbraket{O|\rho}=  \sbra{\CM^{-1} \lr{O}} \mathbb{E} \lrq{ \CC_i \sket{\Pi_{z_i}}},
\end{equation}
where we we also used that $\CM^{-1}$ is self-adjoint. 
This protocol is particularly appealing in practical NISQ-era questions for two reasons.  1.  the projective nature of the channel twirl projects noisy versions of the measurement $\CM_Z$ onto $\CM$ (possibly with adapted coefficients \cite{Chen_2021}) and 2.  a relatively small numbers of samples from the distribution over $(C,  z)$ for a given $\rho$ allows to estimate expectation values of exponentially many well-bounded observables to high confidence \cite{Huang_2020}. Here, the samples $(C, z)$ can be generated in a quantum experiment without yet having decided on the observable $O$.  Due to the design property of the twirl the partial tomographic data obtained this way suffices to estimate selected observables in purely classical post-processing. We refer to the statistical ensemble of samples $(C, z)$ colloquially as the \textit{shadow} of the state.

For applications in quantum computation with bosonic quantum error correction,  such as using the GKP code,  the shadow tomography protocol is particularly interesting since,  typically,  measurements of logical observables can only be carried out indirectly using more naturally accessible measurements such as homodyne detection, heterodyne detection, or photon counting and do not admit a simple and direct physical measurement procedure.  
The projection property (1.) of the logical Clifford-twirl allows us to naturally use more accessible measurements that may be badly aligned with the observables of interest and prescribe how to classical post-process the results to estimate the logical observable at hand. Another
important implication of this property is that classical shadow protocols designed this way are inherently robust to imperfections in the POVM, as the imperfections are projected together with the POVM onto the effective channel and only a potential change of the effective channel parameters needs to be accounted for. This perspective was leveraged in ref.~\cite{Chen_2021} already for discrete variable shadow tomography.
Furthermore,  the effective decomposition of the state \eqref{eq:approx_state} obtained from the shadow protocol is interesting as it may give rise to new representations of states in phase space that capture core logical information.  After discussing how a Clifford channel-twirl can be set up for the GKP code,  we will see how the shadow protocol gives rise to an approximation of GKP states using Gaussian states by logical Clifford twirling a heterodyne measurement. We expect this representation to be particularly useful in the development of new simulation methods for GKP error correction.

\subsection{Displacement twirling}

In this subsection, we turn to discussing twirls that can 
be realized by implementing appropriate displacements in the 
physical Hilbert space. This is a core technical ingredient for the following developments and we discuss its the application of displacement twirling to states and channels.

In ref.~\cite{Conrad_2021}, approximate twirls over groups of displacement operators distributed over lattices $\CL^{\perp}$ associated to the GKP code have been constructed by approximating the uniform measure over the (infinite) lattice $\CL^{\perp}$ via a random walk over a generating set given by the rows of $M^{\perp}=\lr{\bs{\xi}^{\perp\, T}_1,\hdots, \bs{\xi}^{\perp\,  T}_{2n}}^T$.  
Concretely,  we define a random walk from the joint distribution of $N'=2N$ half-steps $\pm \bs{\xi}_i / 2$,  each of which are selected with $1/2$ probability at each step.  Define for $i=1,\hdots, 2n$ the 
associated (discrete) measure \begin{equation}
\mu'_{i}\lr{\bs{x}} =\frac{1}{2}\delta \lr{\bs{x}-\bs{\xi}^{\perp}_i/2}+\frac{1}{2}\delta \lr{\bs{x}+\bs{\xi}^{\perp}_i/2},
\end{equation}
so that we obtain the measure corresponding to $m$ steps of the 
random walk as the $2m$-fold convolution $\mu_i^{(*m)} :=\mu_i'^{(*2m)} $.

\paragraph{State twirling.} 
Applying this twirl to a state,  we obtain that
\begin{align}
\rho &=\int_{\R^2} d\bs{\alpha}\, \rho\lr{\bs{\alpha}} D\lr{\bs{\alpha}}  
\mapsto \int_{\R^2} d\mu_i^{(*m)}\lr{\bs{\gamma}} \, D\lr{\bs{\gamma}} \rho D^{\dagger}\lr{\bs{\gamma}}  
=\int_{\R^2} d\bs{\alpha} \,  \lrq{\nu_i\lr{\bs{\alpha} }}^m \rho\lr{\bs{\alpha}} D\lr{\bs{x}}  \label{eq:state_twirl}
\end{align}
modifies the characteristic function of the state with the $m-$th power of the symplectic Fourier transform of the measure 
\begin{align}
\nu_i\lr{\bs{\alpha}}&= \int_{\R^2} d\mu_i \lr{\bs{\gamma}} e^{-i2\pi \bs{\gamma}^TJ\bs{\alpha}} 
=\cos^2\lr{\pi \lr{\bs{\xi}_i^{\perp}}^TJ\bs{\alpha}}.\label{eq:meas_fourier}
\end{align}
In the limit $m\rightarrow \infty $, this suppresses all contributions $\bs{\alpha}$ except for those in the symplectic dual of $\bs{\xi}_i^{\perp}$.  We define the joint measure over all generators in $M^{\perp}$ to be the joint random walk given by the $2n$-fold convolution
\begin{equation}
\mu_{M^{\perp}} = \mu_1 * \mu_2 *\hdots *\mu_{2n}
\end{equation}
which has  the Fourier transform
\begin{equation}
\nu_{M^{\perp}}\lr{\bs{\alpha}}=\prod_{i=1}^{2n} \cos^2\lr{\pi \lr{\bs{\xi}_i^{\perp}}^TJ\bs{\alpha}}.
\end{equation}
The total effect of this twirl is that in the limit $m\rightarrow \infty$ only logically trivial contributions $\bs{\alpha} \in \CL$ will survive while all other contributions are exponentially suppressed and the state becomes logically fully depolarized.  An alternative view is that in this limit eq.\  \eqref{eq:state_twirl} converges to the group projector of the group generated by the displacement in $\CL^{\perp}$.  The commutant of this group is the stabilizer group generated by displacements in $\CL$.  Non-trivial displacements of $\CL^{\perp}$ are not in this commutant such that it cannot carry logical information.  While twirling a state over $\CL^{\perp}$ does not appear to bear any interesting applications outside deliberate logical depolarization of the state~\footnote{Except perhaps for error mitigation methods where tunable noise is desired as in zero-noise extrapolation \cite{BenjaminZNE,  TemmeZNE}.} 
note that the above outlined method of state twirling is not restricted to using generators in $M^{\perp}$. 
A stabilizer twirl using generator $M$ can equally be used to (approximately) project the state onto one one with a characteristic function supported only on $\CL^{\perp}$.  In conjunction with a twirl over a set of Gaussian unitaries representing a set of logical Clifford gates such as the logical $\hat{H}$ for the square GKP code or $\hat{H}\hat{P}^{\dagger}$, we expect this technique to be useful for the measurement-less preparation 
of \emph{magic states} \cite{PhysRevA.71.022316}, 
analogous to previous proposal for entanglement distillation \cite{Bennett_1996} as well as and the preparation of entangled GKP states analogous to the procedure in refs.~\cite{Werner89, Bennett_1996}.

\begin{figure}
\center
\begin{tikzpicture}[scale=1.1]
\node at (0,0){\includegraphics[width=4cm]{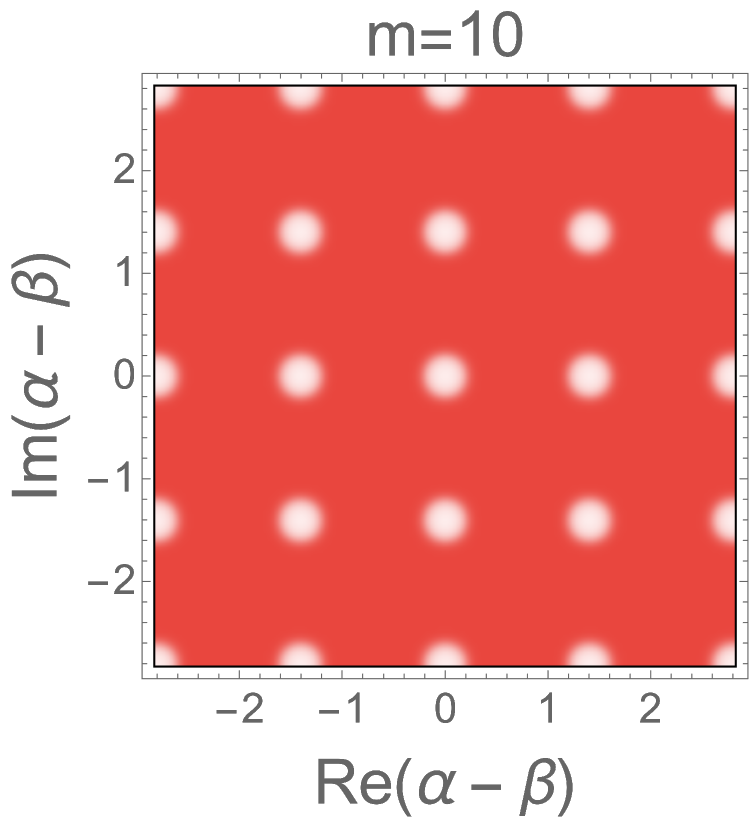}};
\node at (4,0){\includegraphics[width=4cm]{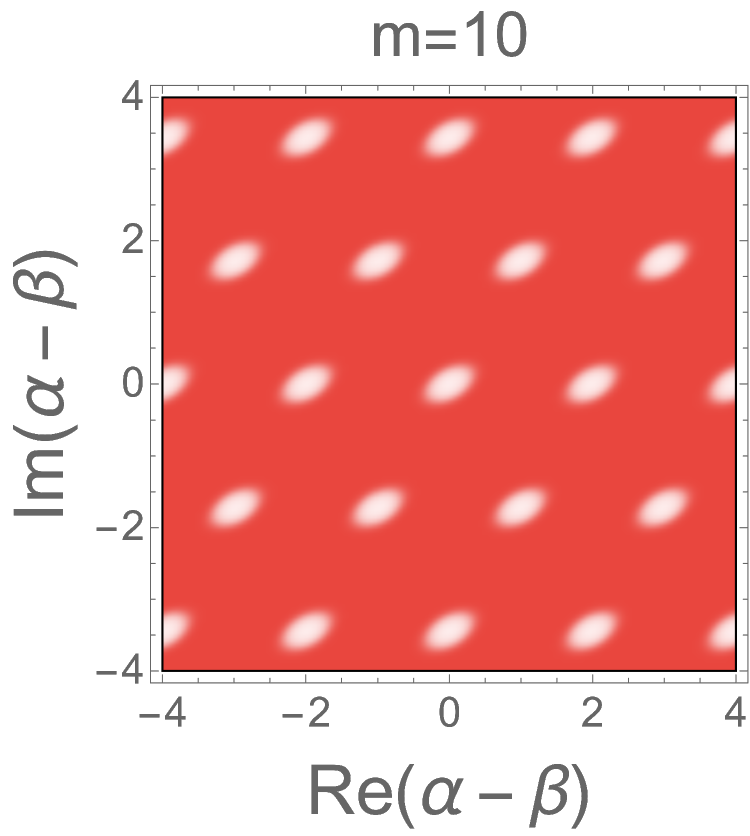}};
\node at (7,0){\includegraphics[width=1.6cm]{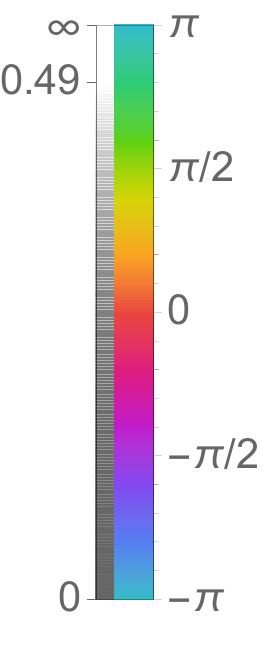}};
\end{tikzpicture}

\caption{Figure for $\nu_{M^{\perp}}$ for square and hexagonal GKP codes.} \label{fig:nuMp}
\end{figure}

\paragraph{Channel twirling.} 
Acting on a quantum channel (see ref.~\cite{Conrad_2021}) 
\begin{equation}
\CC = \int_{\R^{2n}} d\bs{\alpha}\int_{\R^{2n}} d\bs{\beta} \,   c\lr{\bs{\alpha},  \bs{\beta}} D\lr{\bs{\alpha}} \otimes  D^*\lr{\bs{\beta}}
\end{equation}
with chi-function $c\lr{\bs{\alpha},  \bs{\beta}}$,  the $m-$fold displacement channel twirl using our measure $\mu_{M}^{\perp}$ implements the action on the chi-function
\begin{equation}
c\lr{\bs{\alpha},  \bs{\beta}} \mapsto \lrq{\nu_{M^{\perp}}\lr{\bs{\alpha}-\bs{\beta}} }^m c\lr{\bs{\alpha},  \bs{\beta}}.
\end{equation}
Similar to the above,  this channel twirl approximately projects the channel onto a channel where non-stabilizer coherences are suppressed,  i.e.,  contributions in the chi-function $c\lr{\bs{\alpha},  \bs{\beta}}$,  for which $\bs{\alpha}-\bs{\beta} \not\in \CL$ become exponentially suppressed as $m\rightarrow \infty$.  We visualize the factor $\nu_{M^{\perp}}$ for the square-  and hexagonal GKP code 
in fig.~\ref{fig:nuMp}.

For finite strength $m$ of the twirl, the error can be bounded as follows. Let 
\begin{equation}
\overline{\nu}_{\CL}\lr{\Delta} :=\lim_{m\rightarrow \infty} \nu_{M^{\perp}}\lr{\Delta}^m,
\end{equation}
in this limit the function is independent of the choice of dual generating set $M^{\perp}$, which is why the index has been replaced by $\CL$.
For any finite $m$, we have
\begin{align}
    \nu_{M^{\perp}}\lr{\Delta}^m 
    &= \overline{\nu}_{\CL^{\perp}}\lr{\bs{\Delta}} + \nu_{M^{\perp}}\lr{\bs{\Delta} }^m\lrq{1-\overline{\nu}_{\CL^{\perp}}\lr{\bs{\Delta}}}.
\end{align}
To bound the error term, observe that the contribution $\lrq{1-\overline{\nu}_{\CL^{\perp}}\lr{\bs{\Delta}}}$ is only non-zero when $\bs{\Delta} \not\in \CL$. Let $\bs{x}={\rm CVP}\lr{\Delta, \CL}$ be the closest vector in $\CL$ to $\bs{\Delta}$ and let $\bs{\delta}=\bs{\Delta}-\bs{x}$ be the corresponding minimal vector between $\bs{\Delta}$ and the lattice. Assuming $\bs{\delta}$ is small, we can bound each cosine term 
\begin{align}
    \cos^2\lr{\pi \lr{\bs{\xi}_i^{\perp}}^TJ\Delta}
    &= e^{-\pi^2\|\lr{\bs{\xi}_i^{\perp}}\|^2\|\bs{\delta}\|^2}+O\lr{\|\bs{\delta}\|^4},
\end{align}
such that, in total, we have for $\bs{\Delta} \not\in \CL$ close to the lattice
\begin{equation}
    \nu_{M^{\perp}}\lr{\Delta}^m = e^{-\pi^2\|M^{\perp}\|_F^2\|\bs{\delta}\|^2 m }+O\lr{\|\bs{\delta}\|^{4m}},
\end{equation}%
where $\|\cdot\|_F$ denotes the Frobenius norm and $\bs{\delta}=\bs{\Delta}- {\rm CVP}\lr{\bs{\Delta}, \CL}$ is the minimal distance between $\bs{\Delta}$ and $\CL$.

Note that in the above construction we have decided to work with the twirl induced by the described random walk because it yielded a particularly nice analytic form for the characteristic function
\begin{equation}
    \nu_{M^{\perp}}\lr{\bs{x}} = \int_{\R^{2n}} d\mu_{M^{\perp}}\lr{\bs{\gamma}} e^{-i2\pi \bs{\gamma}^TJ\bs{x}}.
\end{equation}
There is no other particularly good 
reason to use this parametrization. In general, 
one may 
also choose an arbitrary regularizer $R\lr{\bs{\gamma}}$ to regularize a uniform distribution over the lattice $\CL^{\perp}$ as \begin{equation}
    d\mu\lr{\bs{\gamma}}=d\bs{\gamma}\,R\lr{\bs{\gamma}} \sum_{\bs{\xi}^{\perp} \in \CL^{\perp}} \delta^{(2n)}\lr{\bs{\gamma} - \bs{\xi}^{\perp}}.
\end{equation}
By the properties of the Fourier transform the result will be provided by the convolution of the asymptotic characteristic with the Fourier-transform of the regularization $\hat{R}$, which will 
be more localized the more homogeneous $R$ is,
\begin{equation}
    \nu\lr{\bs{\Delta}}= \lr{\hat{R} * \overline{\nu}_{\CL}\lr{\bs{\Delta}}}.
\end{equation}

\subsection{Gaussian unitary twirling}

In addition to diagonalizing a state- or channel in the displacement (or Pauli-) basis, the diagonal components can also further be symmetrized. This was discussed previously by twirling over the Clifford group for (qubit-) Pauli-operators. Here we follow the same rational to symmetrize the logical components of a state or channel relative to a prescribed GKP code. It is convenient the logical Clifford group for the GKP code can be implemented using only Gaussian unitary operations, since this leads to a relatively simple analysis of the physical incarnation of the relevant twirl. It will turn out later that some physical measurement channels possess natural symmetries that remedy the necessity to symmetrize them further via a twirl. We will exploit this feature later in the explicit examples using heterodyne measurements and photon-parity measurements.
We implement a random logical Clifford operation as a Gaussian unitary twirl parametrized by a symmetric measure over the symplectic matrices $\mu\lr{S}=\mu\lr{S^{-1}}$,  which implements the map on the displacement twirled chi-function
\begin{align}
\lrq{\nu_{M^{\perp}}\lr{\bs{\alpha}-\bs{\beta}} }^m c\lr{\bs{\alpha},  \bs{\beta}} \nonumber 
&\mapsto \int d\mu\lr{S}  \lrq{\nu_{M^{\perp}}\lr{S^{-1}\lr{\bs{\alpha}-\bs{\beta}}} }^m
c\lr{S^{-1}\bs{\alpha},  S^{-1}\bs{\beta}}  \nonumber\\
&= \int d\mu\lr{S}  \lrq{\nu_{M^{\perp}S^T}\lr{\bs{\alpha}-\bs{\beta}}}^m .
\end{align}
Although such random unitary modifications will never truly project a channel onto one that acts solely within code space,  our goal is to modify it such that for any logical input state and any channel,  decoded logical readout will make it appear as if the twirled channel was a full logically depolarizing.  I.e.,  we target that the projection of the channel onto code space to behave like a logically depolarizing channel.

Note that for the explicit examples we consider here $n=1$ is small enough to simply enumerate the corresponding logical groups $\Sp_2\lr{\Z_d}$ and sample directly from those sets. Nevertheless, we briefly outline how the twirl works in larger systems where the size of $\Sp_2\lr{\Z_d}$ grows super-exponentially. The idea is to choose a generating set $G$ for the symplectic automorphism group $\CG=\Aut^S\lr{\CL^{\perp}}$ and generate random group elements by choosing random sequence from this generating set. It is a result from the theory of random matrices \cite{Varju, DiaconisShashashahani} that a close-to uniform distribution over the target group is obtained by performing at least $k$ steps, once $k\geq |G|\mathrm{diam}^2\lr{\CG,  \,G}$, where $|G|$ is the size of the generating set and the group diameter $\mathrm{diam}\lr{\CG,  \,G}$ expresses the minimal number of generators needed to express any element in the group $\CG$.

For $\CG=\Sp_{2n}\lr{\Z_d}$ with generating set given by elementary generalized Hadamard, phase and CNOT gates  (see appendix), which in their real representation are given by symplectic transvections, this evaluates to a bound $k\geq O\lr{d^2n^6}$.   
This bound can be slightly improved to $k\geq O\lr{d n^6}$ when adding the $r=0, \hdots, d-1$ powers of the elementary gates.
We hence obtain that a logical twirl over the full (non-trivial) logical Clifford can be realized by random sequences of $k\geq O\lr{d^2n^6}$ elementary symplectic transvections.

\section{GKP logical shadows}\label{sec:GKPShadows}
In the previous section we have discussed how a general CV channel $L$ can be twirled using displacement operators and Gaussian unitary operations such that it effectively yields a logical depolarizing channel. 
We here turn to showing how these insights and the
established machinery can be turned
into a scheme devising GKP logical shadows for the efficient measurement of expectation values of observables on the logical level.
The key to this analysis is to develop an understanding of how the physical twirl manifests itself on the logical level, where the conversion from a physical representation of the state to its logical information content is facilitated by a decoder.

\subsection{From physical to logical twirls}

In the following we discuss the how the physical structure of a twirl manifests on the logical level. Understanding this connection allows us to build logical shadow tomography protocols based on physical measurements statistics. This approach is more versatile and effective than projecting a physically reconstructed state onto the code space of a chosen code, an approach chosen in ref.~\cite{hu2022logicalshadowtomographyefficient}, but not suitable for the GKP code.

For any physical channel $\CC$,  denote by $\CC^{\tau}$ the channel twirled using logical displacements from $\CL^{\perp}$ and Gaussian unitaries represented by $\Aut^S\lr{\CL^{\perp}}$ as discussed earlier.
We define a \textit{decoder} $\Dec$ to be an surjective map from physical states to logical,  error-free states $\Dec\sket{\rho}=\sket{\overline{\rho}}$.  The decoder is required to commute with noiseless logical channels $\Dec\CC=\CC\Dec$,  where $\CC$ are logical Clifford operations (appropriately represented by a Gaussian unitary channel) and reduce error channels $\CR$ compactly supported on displacements from the Voronoi cell 
\begin{equation}
\mathcal{V}\lr{\CL^{\perp}}:=\lrc{\bs{x}\in \R^{2n},  \|\bs{x}\|\leq \|\bs{x}-\bs{\xi}^{\perp}\|\; \forall \bs{\xi}^{\perp} \in \CL^{\perp}}
\end{equation}
to the logical identity channel $\Pi=\Pi_{\CL}\otimes\Pi_{\CL}^*$,  $\Pi\CR= \Pi$,  where $\Pi_{\CL}$ is the code space projector. 

Applying the decoder to the twirled channel,  we obtain for any 
input state $\rho$
\begin{align}
 \Dec\CC^{\tau} \sket{\rho}&=\alpha \Dec\sket{\rho} + \beta \sket{\Pi_{\CL} }
 \nonumber
 \\
 &=\alpha\sket{\overline{\rho}}+\beta \sket{\Pi_{\CL} }:=\widetilde{\CM}\sket{\overline{\rho}},\label{eq:DTwirled}
\end{align}
that the decoded twirled channel effectively behaves like a logically depolarizing channel. This decomposition results from the fact that for any logical Clifford operation we have that $\Dec\CC=\CC\Dec$, such that the decoder in particular also commutes with the logical twirl. On the (decoded) logical level it is already known that a Clifford twirl projects on a channel of the above form of a depolarizing channel.
The depolarizing channel is invertible as $\widetilde{M}^{-1}\lr{X}=\lr{X-\beta\Pi_{\CL}}/\alpha$ as long as $\alpha\neq 0$.  
Let $\CC=\sum_i\sketbra{\Pi_i}$ be the channel corresponding to a projective POVM (see also a formulation using the theory of measurement frames~\cite{Innocenti}). 
Equation~\eqref{eq:DTwirled} takes the form
\begin{equation}
\widetilde{\CM}\sket{\overline{\rho}}= \Dec  \sum_i\int_{\rm GU\lr{n; \CL}} d\mu\lr{U} \,  \CU\sket{\Pi_i} \sbraket{\Pi_i|\CU^{\dagger}|\rho},
\end{equation}
where the unitaries are drawn from a distribution over Gaussian unitary operations representing logical Clifford gates obtained from concatenating the displacement- and non-trivial logical Clifford ${\rm GU\lr{n; \CL}}$, as discussed in the previous sections. Since $\widetilde{\CM}$ commutes as logical channel with $\Dec$,  given a physical state $\sket{\rho}$,  we can obtain the expectation value of any logical observable $O$ via the following shadow protocol.
\begin{enumerate}
\item Sample a set of $\lrc{\CU_i}_{i=1}^N$ from ${\rm GU\lr{n; \CL}}$ and perform the POVM for each sample.  The probabilities of obtaining outcome pointers $\sket{\Pi_i}$ are each given by the Born probability $\sbraket{\Pi_i| \CU^{\dagger}_i | \rho}$.
\item Reconstruct the shadow as empirical expectation value over the output states \begin{equation}\sket{S}=N^{-1}\sum_{i=1}^N  \CU_i \sket{\Pi_i}.  \label{eq:shadowstate}\end{equation}
\item Compute the logical expectation value \begin{equation}\sbraket{O|\widetilde{M}^{-1}|S}=\sbraket{\widetilde{M}^{-1}\lr{O}|S}.\end{equation}
\end{enumerate}
Since 
\begin{equation}
\Dec\mathbb{E}\lrq{ \widetilde{M}^{-1} \CU \sket{\Pi}} = \Dec\sket{\rho},  \label{eq:Drho_shad}
\end{equation}
for sufficiently large sample size the empirical mean state converges to the true mean state and the procedure recovers the expectation value of the logical observable as if it was measured directly after implementing a suitable decoder, that is, 
the estimator 
\begin{equation}
\tilde{o}=\sbraket{\widetilde{M}^{-1}\lr{O} | \CU| \Pi}
\end{equation}
inherits the mean $\langle \tilde{o}\rangle=\sbraket{O|\Dec|\rho}$.

It is key to the protocol, however, that not only the mean is correctly recovered, but that also the variance is small: on this logical level the estimation essentially does not differ from the typical qubit scenario as already detailed in the construction provided by Huang,  Kueng and Preskill \cite{Huang_2020} and their performance guarantees apply,  in that a small number of samples suffices to reproduce the expectation value of many observables with high confidence.

\begin{them}[HKP \cite{Huang_2020}]
A collection of $NK$ samples $\lrc{\lr{\CU_i , \Pi_i }}_{i=1}^{NK}$ produced via the above protocol from a CV state $\rho$ suffice to estimate logical expectation values on observables $O_i,\; i=1, \hdots , M$  via median of means prediction up to $\epsilon$ additive error provided that
\begin{equation}
K=2\log\lr{2M/\delta},\; N=\frac{34}{\epsilon^2} \max_i \biggl\|O_i-\frac{\Tr\lrq{O_i}}{2^n} I \biggr\|^2_{\rm shadow}
\end{equation}
with probability at least $1-\delta$. 
\end{them}
We refer to ref.~\cite{Huang_2020} for the proof and the definition of the shadow norm, 
for which ref.\ \cite{Huang_2020} 
has also provided the upper bound 
\begin{equation}
\biggl\|O_i-\frac{\Tr\lrq{O_i}}{2^n} I \biggr\|^2_{\rm shadow}\leq 3\Tr\lrq{O_i^2}.
\end{equation}
Median of means prediction 
is carried out by dividing the set of $NK$ samples into $K$ batches of $N$ samples each,  for each of which the arithmetic mean is computed  and taking the median over the batches
\begin{equation}
\tilde{o}^{\rm est} = {\rm median} \lrc{ N^{-1}\sum_i \tilde{o}_{jN+i} }_{j=1,\hdots
, K}.
\end{equation}
In particular,  this also implies that the reconstructed state \eqref{eq:shadowstate} yields a good representation of the decoded state $\Dec\sket{\rho}$ in that it reproduces the expectation value of many low-rank observables with only small additive error.  By paying an extra cost in sample overhead,  we can use the shadow to obtain a full representation of the (finite dimensional)
state.

\begin{them}[Full representation of the 
state]\label{them:GKPshadow}
Let $\CL \subset \R^{2n}$ denote the lattice corresponding to a scaled GKP code on $n$ modes that encodes $d^n$ logical dimension and let $\mu$ denote a measure over elements $\Aut^S\lr{\CL^{\perp}}$ forming a logical Clifford $2-$design and let $\CM=\SumInt dz \sketbra{\Pi\lr{z}}$ denote a physical projective POVM. Let $\sket{\rho}$ be an arbitrary state vector on the $n$-mode Hilbert space. The state vector
\begin{equation}
\sket{S}=N^{-1} \sum_{i=1}^N \CU_{S,  i}\sket{\Pi\lr{z_i}}
\end{equation}
produced by sampling $N$ Gaussian unitary operations via the measure $\mu$ and measurement outcomes from the Born distribution $z_i \sim \sbraket{\Pi\lr{z}|\CU_{S,  i}^{\dagger}|\rho}$ approximates the logical value of the state in Hilbert-Schmidt distance

\begin{equation}
d_{\rm HS}\lr{\Dec \sket{\rho} ,  \Dec \sket{S}} \leq \delta_{\rm HS}^2
\end{equation}
with probability at least $1-\delta$ for 
\begin{equation}
N\geq \frac{2d^{2n}}{\alpha^2 \delta_{\rm HS}^2} \lrq{\ln\lr{\frac{2}{\delta}} + 2n \ln\lr{d}},
\end{equation}
where $\alpha$ is determined by the commutation of the twirled 
POVM with the decoder
\begin{align}
 \Dec\CM^{\tau} &=\alpha \Dec + \beta \sket{\Pi_{\CL} }.
\end{align}
In particular, 
we also have
\begin{equation}
\|\Dec \sket{\rho}-  \Dec \sket{S}\|_1 \leq d^{\frac{n}{2}}\frac{\delta_{\rm HS}}{2} . 
\end{equation}
\end{them}
See appendix \ref{app:GKPshadow_proof} for the proof of this statement. Although the sample overhead derived here contains the extra dimensional factor $O\lr{d^{2n}}$ and is therefore
rather large compared to the usual situations considered in ref.~\cite{Huang_2020}, this statement allows us to apply the shadow tomography toolbox to derive simple representations of bosonic states with the right information content on selected (GKP) subspaces. The extensive overhead, contrasting the result of ref.~\cite{Huang_2020}, here essentially stems from ``abusing'' the shadow tomography routine to perform full (logical) state tomography, requiring the well-estimation of a complete number of observables scaling with the system size. While this overhead may seem daunting, estimating expectation values of observables is not our primary goal. Instead, we emphasize that the shadow tomography protocol allows for a decomposition of the input state into a mixture of pointer states that reproduces the logical information of the encoded state. 
The sample complexity bounds imply an upper bound on the number of components in this mixture. This bound is generically loose, but the fact that it can be stated and derived using the tools used for shadow tomography we believe is nontrivial. While decompositions may also be formalized using the language of measurement frames \cite{Innocenti}, the presented toolbox allows to quantify the strength of the decomposition in a flexibly in the chosen POVM We exemplify its utility using the heterodyne- and photon-click 
POVM that resolve the presence and absence of photons.

\subsection{Gaussian decomposition of the hexagonal GKP code from heterodyne measurements}

An interesting use case of our protocol with practical applications is obtained by considering heterodyne measurements paired with the hexagonal GKP code. Heterodyne measurements are such that the measurement pointer states are Gaussian states. Furthermore, the measurement channel is rotationally symmetric, which will turn out to skip the need to apply non-trivial Gaussian operations in the implementation of our protocol when paired with the hexagonal GKP code. After developing the basic ingredients to this protocol, we will later discuss how it can be applied to simulate the forward-propagation of general quantum states through CV circuits.

In our convention for displacement operators,  we define generalized coherent state vectors as $\ket{\bs{\alpha}}=D\lr{\bs{\alpha}}\ket{0}$,  where $\ket{0}$ denotes a $n$-mode vacuum state where we have
\begin{align}
\ketbra{\bs{\alpha}}&=\int_{\R^{2n}} d\bs{\beta}\,  \Tr\lrq{D^{\dagger}\lr{\bs{\beta}} \ketbra{\bs{\alpha}}} D\lr{\bs{\beta}} \\
&=\int_{\R^{2n}} d\bs{\beta}\,   e^{-\frac{\pi}{2} \bs{\beta}^T\bs{\beta}-i2\pi\bs{\alpha}^TJ\bs{\beta}}D\lr{\bs{\beta}} ,
\nonumber
\end{align}
such that a resolution of the identity is given by
\begin{equation}
\int_{\R^{2n}}d\bs{\alpha}\, \ketbra{\bs{\alpha}} = I.
\end{equation}
In this sense, the coherent states constitue an overcomplete resolution of the identity.
We also have
\begin{equation}
\braket{\bs{\beta}|\bs{\alpha}}=e^{-\frac{\pi}{2}\lr{\|\bs{\alpha}-\bs{\beta}\|^2+i2\bs{\alpha}^TJ\bs{\beta}}}.
\end{equation}
Generalized heterodyne measurements are interferometric quantum optical measurements
that are known to in effect implement a projective POVM described by the quantum channel
\begin{align}
\CC&=  \int_{\R^{2n}} d\bs{\alpha}  \ketbra{\bs{\alpha}} \otimes  \ketbra{\bs{\alpha}}^* \\
&=\int_{\R^{2n}} d\bs{\beta} e^{-\pi \bs{\beta}^T\bs{\beta}} D\lr{\bs{\beta}} \otimes D\lr{\bs{\beta}}^*,
\nonumber
\end{align}
equivalent to a Gaussian displacement channel where displacements of amplitude $\bs{\beta}$ occur with probability density $ e^{-\pi \bs{\beta}^T\bs{\beta}}$. The above equation is obtained by expanding each coherent state $\ketbra{\bs{\alpha}}$ in displacement operators and executing the integral over the variable $\bs{\alpha}$.
In this form it becomes clear why this POVM is special.  1.  it is already diagonal in displacement operators,  such that a displacement twirl has no effect and 2.  the chi-function only depends on the Euclidean length $\|\bs{\beta}\| $ of the corresponding displacement amplitudes,  such that the channel is also invariant under Gaussian unitary twirls realized via orthogonal symplectic transformations of $\bs{\beta}$. 

For the single mode hexagonal GKP code we have found that a Clifford $2$-design is given by the set of rotations $R_{\frac{2\pi}{3}}$,  which are such orthogonal symplectic transformations.  Hence this is an example where the twirls have no effect.  
We can define a sequence of \textit{Voronoi-``shells''}
\begin{align}
\mathcal{V}^0&=\mathcal{V}\lr{\CL^{\perp}},\\
\mathcal{V}^k&=\mathcal{V}\lr{(2k+1)\CL^{\perp}}\setminus \mathcal{V}^{k-1}, \; k\in \N_0,\\
\bigcup_{k=0}^{\infty}\mathcal{V}^k&=\R^{2n},
\end{align}
where each shell with $m=0\, \mod d$ contains logically trivial displacement amplitudes.  
In particular  $ \forall \bs{x}\in \mathcal{V}\lr{\CL^{\perp}}$ our decoder maps
\begin{equation}
\Dec \CD\lr{\bs{x}} = \Dec.
\end{equation}
For $d=2$, 
we can 
compute
\begin{align}
\Dec \CC &=\Dec\sum_{k=0}^{\infty} \int_{\mathcal{V}^k}d\bs{\alpha}\, e^{-\pi\bs{\alpha}^T\bs{\alpha}} \\
\nonumber
&=(p_0-p_1)\Dec + 2p_1\Pi_{\CL}, \\ \hspace{.5cm}
p_i&=\sum_{k=0}^{\infty} \int_{\mathcal{V}^{2k+i}}d\bs{\alpha}\, e^{-\pi\bs{\alpha}^T\bs{\alpha}}
\end{align}
where we have used that each displacement in an even shell is removed by the decoder and each displacement in an odd shell is equally likely attributed to a logical Pauli $X-,\, Y-,$ or $Z-$ displacement.  Here we have 
that the associated logical depolarizing channel $\widetilde{M}=(p_0-p_1)\Dec + 2p_1\Pi_{\CL}$ is invertible as long as $p_0\neq p_1$.

Since we have $p_0+p_1=1$,  this is can be verified by showing that $p_0\neq 1/2$. 
We have $\mathcal{V}^0\supseteq \CB^2\lr{\rho\lr{\CL^{\perp}}}$ and for $k\geq 1$
\begin{equation}
\mathcal{V}^k \subseteq \CB^2\lr{\lr{2k+1}\mu\lr{\CL^{\perp}}} \setminus \CB^2\lr{\lr{2k-1}\rho\lr{\CL^{\perp}}},
\end{equation}
where $\rho\lr{L}=\lambda_1\lr{L}/2$ denotes the packing radius of the lattice $L$ and $\mu\lr{L}$ denotes the covering radius,  which for the hexagonal lattice is given by $\mu\lr{A_2}=\lambda_1\lr{A_2}/\sqrt{3}$ where $\lambda_1 \lr{A_2}=\sqrt{2/\sqrt{3}}$.
Using $\lambda_1\lr{\CL^{\perp}}=\lambda_1\lr{A_2}/\sqrt{2}$ we can thus bound
\begin{align}
p_0&=\sum_{k=0}^{\infty} \int_{\mathcal{V}^{2k}} e^{-\pi \|\bs{\alpha}\|^2} \leq  2\pi\int_0^{\mu\lr{\CL^{\perp}}} r e^{-\pi r^2}  \\
\nonumber
&\hspace{1cm}+  2\pi \sum_{k=1}^{\infty} \int_{(4k-1)\lambda_1\lr{\CL^{\perp}}/\sqrt{3}}^{(4k+1)\lambda_1\lr{\CL^{\perp}}/2} dr\, r e^{-\pi r^2}  \\
\nonumber
&=\lrq{1-e^{-\frac{\pi}{3\sqrt{3}}}} \\ 
\nonumber
&\hspace{.5cm} + \sum_{k=1}^{\infty} \lr{e^{-\frac{\pi}{3\sqrt{3}}(4k-1)^2\lambda_1^2} - e^{-\frac{\pi}{4\sqrt{3}}(4k+1)^2\lambda_1^2}}\\
&\approx 0.455\hdots \, .
\nonumber
\end{align}
This yields a bound $|\alpha|=|p_0-p_1|=|2p_0-1| \geq 0.09$.
We find that while the effective channel $\widetilde{M}$ is in fact invertible,  the logical information retained in the depolarized state is rather minuscule as indicated by the value $|\alpha|=|p_0-p_1|=|2p_0-1| \geq 0.09$.

\subsection{Decoding}\label{sec:decoding} We briefly comment on the classical post-processing required to execute this protocol where, w.l.o.g, we focus on the logical Pauli-Z measurement outcomes. Given a sample corresponding to a pure Gaussian state $\rho_{G}$ for a GKP code with lattice $\CL$, there exist a Gaussian unitary transformation implementing a symplectic transformation $S$ such that $S\CL = \bigoplus_{i=1}^n \CL_{i}$, where each $\CL_{i}$ describes a local square single mode GKP code encoding $d_i$ dimensions. We denote the Gaussian state resulting from this transformation by $\rho_{G'}$. The decoding map then effectively implements
\begin{equation}
    \overline{\rho_Z}=\bigotimes_{i=1}^n \mathrm{dec}_i\lr{\rho_{G'}},
\end{equation}
where with $k_i=1,..,d_i-1$
\begin{equation}
    \mathrm{dec}_i\lr{\rho_{G'}} = \sum_{k_i=0}^{d_i-1} \int_{S_{k_i, i}} \Tr\lrq{\rho_{G'} \ketbra{x_i}_q} dx_i\; \ketbra{\overline{k_i}}\label{eq:dec_i}
\end{equation}
represents the ``binned homodyne detection'' and 
\begin{equation}
    S_{k_i, i} = \bigcup_{s\in \Z} \lrq{-\frac{1}{2}\sqrt{\frac{2\pi}{d_i}}+\sqrt{\frac{2\pi}{d_i}}k_i + \sqrt{2\pi d_i}s,\, \frac{1}{2}\sqrt{\frac{2\pi}{d_i}}+\sqrt{\frac{2\pi}{d_i}}k_i + \sqrt{2\pi d_i}s}.\label{eq:bins}
\end{equation}
As $\rho_{G'}$ is Gaussian, this decoding step can be approximated using suitably large truncation $|s|\leq \overline{s}$, which can be estimated to be on the scale of $\mathtt{sq}\lr{S}=\sqrt{\lambda_{\rm max}\lr{S^TS}} )$. 
Note that as
\begin{equation}
    \bigcup_{i=0}^{d_i-1}S_{k_i, i} = \R\; \forall i=1,..,n,
\end{equation}
The single mode decoder in eq.~\eqref{eq:dec_i} should be approximated by examining the relative magnitude over different $k_i$'s of the integral sequence. With each mode being decoded separately, the overall cost in postprocessing can be estimated to scale as $O\lr{n \mathtt{sq}\lr{S}}$.

We do not examine this question further in this work, and remark that the information theoretic decomposition into Gaussian states, and their subsequence use in simulation are not hindered by a potentially costly decoding step.

\subsection{Decomposing GKP states from photon click detectors}

Another practically relevant measurement channel is provided by photon click detectors~\cite{Cova2004}.
Photon click detectors have dichotomic outcomes and distinguish the presence from the absence of photons.
The relevant POVM is given by
\begin{align}
\CC&=\Pi_0\otimes \Pi_0^* + \Pi_{1} \otimes \Pi_1^*, \\
\Pi_0&=\ketbra{0},\\
\Pi_{1}&=I-\ketbra{0},
\end{align}
with chi-function 
\begin{align}
c\lr{\bs{\alpha}, \bs{\beta}}&=e^{-\frac{\pi}{2} \|\bs{\alpha}\|^2-\frac{\pi}{2} \|\bs{\beta}\|^2} \\ 
&\hspace{.5cm}
+\lr{\delta\lr{\bs{\alpha}}- e^{-\frac{\pi}{2} \|\bs{\alpha}\|^2}}\lr{\delta\lr{\bs{\beta}}- e^{-\frac{\pi}{2} \|\bs{\beta}\|^2}}.\nonumber
\end{align}
As before, this channel is rotationally invariant, such that the only non-trivial component of the twirl is given by its diagonalizing action on the chi function. Note that the wave function collapse induced by projective measurement of a generating set of GKP stabilizers is equivalent a full stabilizer displacement twirl (with $m\rightarrow\infty$). Ignoring stabilizer coherences, 
the channel thus takes the form
\begin{align}
 \CC^{\tau} \lr{\rho} &=   \int d^2\bs{\alpha} f\lr{\bs{\alpha}} D\lr{\bs{\alpha}} \rho D^{\dagger}\lr{\bs{\alpha}}, \\
 f\lr{\bs{\alpha}}&= \sum_{\bs{\xi} \in \CL} c\lr{\bs{\alpha},\bs{\alpha}+\bs{\xi}}e^{-i\pi\bs{\alpha}^TJ\bs{\xi}}.
\end{align}
Commuted through a decoder, the effective depolarizing channel obtains the coefficients
\begin{align}
    \alpha&=1-2\vartheta\lr{0 \, | \, iG_{A_2}}+I_1
    ,
    \\
\alpha+\beta &= 1-2\vartheta\lr{0 \, | \, iG_{A_2}}+I_2,
\end{align}
where we have defined the Riemann theta function as
\begin{equation}
\vartheta\lr{\bs{z}\, | \, F}=\sum_{\bs{m}\in \Z^{2n}} e^{i2\pi\lr{\frac{1}{2}\bs{m}^TF\bs{m}+\bs{m}^T\bs{z} }}
\end{equation}
and 
\begin{align}
    I_1&=\sqrt{2} \int_{\mathcal{V}\lr{A_2}}d\bs{\alpha'} e^{-\frac{\pi}{2}\|\bs{\alpha'}\|^2} \vartheta \lr{M_{A_2}\lr{iI+J}\bs{\alpha'}| i G_{A_2} }\nonumber\\
&\hspace{2cm}
\times\vartheta \lr{M_{A_2}\lr{iI-J}\bs{\alpha'}| i G_{A_2} }
,\\
    I_2 &= 2\sqrt{2} \int_{\mathcal{V}\lr{A_2}}d\bs{\alpha'} e^{-2\pi\|\bs{\alpha'}\|^2} \vartheta\lr{M_{A_2}\lr{iI+J}\bs{\alpha'}| i G_{A_2} } \nonumber\\
    &\hspace{2cm}
    \times
    \vartheta \lr{M_{A_2}\lr{iI-J}\bs{\alpha'}| i G_{A_2} },
\end{align}
where $G_{A_2}=M_{A_2}M_{A_2}^T$ is the Euclidean Gram matrix for the symplectic basis of the $A_2$ lattice.
By numerical integration and approximating the Riemann theta function with one evaluated using a truncated sum $\bs{m}\in \lrc{-t,\hdots ,t}^{2n}$ with $t$ chosen large enough for convergence of the output values, we evaluate $\vartheta\lr{0 \, | \,i G_{A_2}}=1.1596$, $I_1=1.493$, $I_2=1.64$ to obtain the estimates $\alpha=0.32$ and $\beta=0.15$. Similar to the previous section, this shows that a representation of the logical content of a single mode GKP code relative to the hexagonal GKP code can be obtained from a finite number of samples from random displaced photon click detectors, of which the number is bounded by theorem \ref{them:GKPshadow}. As a final remark we mention that potential decoding of the pointer states can proceed analogous as discussed in the previous section.

In this section, 
we have adapted the (HKP classical) shadow tomography protocol introduced in 
ref.~\cite{Huang_2020} to operate on the logical degrees of freedom of quantum information encoded via a GKP code in a continuous variable system and considered two of the best-behaved POVMs given by heterodyne detection and a photon-click detector and derived the constant $\alpha$ to show the invertibility of the effective depolarizing channel. While the full analysis is more complex for more general POVMs, such as photon counting, our approach is sufficiently general to be generalized to such situations.

\subsection{Applications to classical simulation}

In this section, we
briefly speculate about applications of the logical shadow tomography schemes developed here. The main development of the preceding sections has been that one can view the GKP logical shadow protocol as a physical black box protocol that converts a given physical state via application of appropriate random Gaussian unitaries into a convex combination of Gaussian states when using heterodyne detection as the underlying POVM. Since Gaussian states are efficiently described by mean $\bs{\overline{x}}\in \R^{2n}$ and variance $V=S^TS\in \R^{2n \times 2n}$, each part of the decomposition offers an efficient classical description and transforms in a simple manner under application of Gaussian channels \cite{Weedbrook_2012}.
Similar decompositions have been utilized, e.g., 
in ref.~\cite{Bourassa_2021} to develop classical simulation methods for CV states by first decomposing them into a linear combination over Gaussian states, analogous to simulation via stabilizer state decompositions found in the qubit-based quantum computing literature, see, e.g., ref.~\cite{Bravyi2019simulationofquantum} and references therein. 
The practical drawback of the previously 
presented methods, however, is that an analytic 
description of the initial state must be known \textit{a priori}, on which basis the decomposition then proceeds and no  bound has been derived on the number of parts necessary to obtain a good approximation of the state. By focusing on reproducing underlying logical properties of the input state and using statistical methods known from classical shadow tomography, the tools developed here yield an experimental method that converts a given physical state into a convex combination of Gaussian states, corresponding to the samples obtained in the protocol and our bounds yield upper bounds on the number of components (samples) necessary to achieve a good approximation to the input state on the logical level. In the development of practical quantum computers based on GKP encoded logic, one may hence apply this protocol to convert experimentally realizable (multi-mode) GKP states into a classical description of Gaussian state components, which can then further be used to assess their performance in algorithms or under noise. While the overhead is not expected to scale favorably in the number of non-Gaussian channels, one may also imagine a recursive version of this procedure to simulate non-Gaussian evolutions while using the logical shadow protocol to repeatedly convert the mid-simulation states into a convex combination of Gaussian states. The protocol outlined here is only one of many possible uses of our scheme and highlights its potential for practical applications.

\subsection{Random Wigner tomography}

The final measurement channel we consider to exemplify our protocol is photon parity measurements. This measurement channel is practically widely implemented and relevant for its application to Wigner tomography \cite{He2024} as we will further discuss below. It will turn out that logical GKP shadows can be understood as a randomized Wigner tomography scheme, where the Wigner function of the input state is only evaluated at random points chosen from a distribution close to the symplectic dual lattice $\CL^{\perp}$ of a GKP code. This yields a protocol which estimates the logical shadow for any state close to a GKP code state but is not limited to such states. For states not close to GKP code states, it merely reproduces a logical shadow of a state \textit{if it was interpreted} as a code state. Later, in sec.~\ref{sec:chasing_shadows}, we consider this protocol relative to random choices of GKP codes and quantify how such protocol yields a shadow for the full physical input state.

The parity operator $\hat{\pi}=e^{-i\pi \hat{N}}$, with $\hat{N}=\sum_{i=1}^n, \hat{n}_i$ has an interesting decomposition into displacement operators with constant characteristic function
\begin{align}
    \Tr\lrq{D^{\dagger}\lr{\bs{\beta}} \hat{\pi}}&=\int_{\R^{2n}}d\bs{\alpha}\, \braket{\bs{\alpha}| D^{\dagger}\lr{\bs{\beta}}|-\bs{\alpha}} \\ 
    &=\int_{\R^{2n}}d\bs{\alpha}\, e^{-2\pi \bs{\alpha}^T\bs{\alpha}} =2^{-n}.
    \nonumber
\end{align}
The fact that this characteristic function is constant has the effect that expectation values of displaced parity measurements yield
\begin{align}
    \langle D\lr{\bs{x}}\hat{\pi}D^{\dagger}\lr{\bs{x}} \rangle 
    &= 2^{-n} \int_{\R^{2n}} d\bs{\beta}\, e^{-i2\pi \bs{x}^T J \bs{\beta}} \Tr\lrq{D\lr{\bs{\beta}} \rho} \nonumber \\
    &= 2^{-n} W_{\rho}\lr{\bs{x}}. 
\end{align}
Up to the rescaling $2^{-n}$  this is precisely the Wigner function we have encountered in eq.\ \eqref{eq:Wigner}. This basic insight has been used consistently in quantum optics for the purposes of tomographic recovery \cite{PhysRevLett.76.4344}.
The chi-function associated to the measurement channel
\begin{align}
    c\lr{\bs{\alpha}, \bs{\beta}} 
    &=\frac{1}{2}\lrq{\delta\lr{\bs{\alpha}}\delta\lr{\bs{\beta}}+2^{-2n}} \label{eq:c_Wigner}
\end{align}
is invariant under symplectic transformations $S\in \Sp_{2n}\lr{\R}:\; c\lr{S\bs{\alpha}, S\bs{\beta}}=c\lr{\bs{\alpha}, \bs{\beta}}$:  the only non-trivial component a logical Clifford twirl can have is provided by the displacement twirl over displacements in $\CL^{\perp}$.  Due to the connection of this POVM to Wigner tomography, each displacement sampled in the displacement twirl $\mu_{\CL^{\perp}}\lr{  \bs{\gamma}} $ corresponds to choosing to estimate the Wigner function of the input state $\rho$ at the random point $\bs{\gamma}\in \CL^{\perp}$. Operationally, the protocol works by 1. sampling a displacement vector from the distribution specified in the twirl and 2. measuring the photon parity  a constant number of times to obtain the expectation value of the displaced photon parity operator or, equivalently, the probabilities the obtain a $\pm 1$ outcome in the parity measurements. The expectation value of that displaced parity operator is exactly the value of the Wigner function at the sampled displacement vector and can be obtained with fixed target precision at a constant number of samples for each displacement vector $\bs{\gamma}\in \CL^{\perp}$ sampled in the twirl. For clarity of presentation we suppress this constant in our analysis. 
Similar as before, a displacement twirled photon parity measurements -- let us call them \textit{Wigner shadows} -- can also be understood to behave like a logical depolarizing channel. In contrast to the cases we have considered before, the characteristic function in eq.\ \eqref{eq:c_Wigner} contains a constant that would yield a diverging contribution when summing over all stabilizer-equivalent coherences in the channel, which forces us to make a more refined model of the displacement twirl. 
We consider the measure $d\mu\lr{\bs{\gamma}}=p_{\sigma^2}\lr{\bs{\gamma};\CL^{\perp}} d\bs{\gamma}$, with probability density
\begin{align}
    p_{\sigma^2}\lr{\bs{\gamma};\CL^{\perp}}&={N_{\CL^{\perp}}\lr{\sigma}}^{-1} \sum_{\bs{\xi}^{\perp} \in \CL^{\perp}} e^{-\frac{\sigma^2}{2}\|\bs{\xi}^{\perp}\|^2-\frac{1}{2\sigma^2} \| \bs{\gamma}- \bs{\xi}^{\perp}\|^2},\\
    N_{\CL^{\perp}}\lr{\sigma}&=(2\pi \sigma^2)^n\, \Theta_{\CL^{\perp}}\lr{i\frac{\sigma^2}{4\pi}},
\end{align}
where we have introduced the lattice theta constant \cite{ConwaySloane, Conrad_2022} $\Theta_{L}\lr{z}=\sum_{\bs{x}\in L} e^{i2\pi z\|\bs{x}\|^2}$ to express the normalization factor.

The characteristic function of this distribution is given by
\begin{align}
    \nu_{\sigma^2}\lr{\bs{\Delta}}&= \int_{\R^{2n}} d\mu\lr{\bs{\gamma}}\, e^{-i2\pi \bs{\gamma}^TJ\bs{\Delta}}
    \\
    &=N_{\CL^{\perp}}\lr{\sigma}\sum_{\bs{\xi}^{\perp} \in \CL^{\perp}} e^{-\frac{\sigma^2}{2}\|\bs{\xi}^{\perp}\|^2-i2\pi \lr{\bs{\xi}^{\perp}}^TJ\bs{\Delta} } 
    \int_{\R^{2n}} d\bs{\gamma}\, e^{-\frac{1}{2\sigma^2}\bs{\gamma}^T\bs{\gamma}-i2\pi \bs{\gamma}^TJ\bs{\Delta}},
    \nonumber
\end{align}
and we can evaluate this line using the 
Poisson summation formula (see 
ref.~\cite{Conrad_2022} for a formulation 
tailored to symplectic inner 
products and sums over lattices) and Gaussian integration to obtain
\begin{align}
    \nu_{\sigma^2}\lr{\bs{\Delta}}&= c^{-1} \sum_{\bs{\xi} \in \CL} e^{-2\pi^2\sigma^2 \|\bs{\Delta}\|^2}e^{-\frac{2\pi^2}{\sigma^2} \|\bs{\Delta}-\bs{\xi}\|^2}, \\
    c&=\det\lr{\CL^{\perp}} \sigma^{2n} \Theta_{\CL^{\perp}}\lr{i\sigma^2/4\pi}.
\end{align}
This distribution is in particular normalizable, with $\int d\bs{\Delta}\, \nu\lr{\bs{\Delta}} =1$.
This allows us to compute the logical fidelity of the depolarizing channel
\begin{align}
    1-p&\geq\int_{\mathcal{V}\lr{\CL^{\perp}}} d\bs{\alpha}\, \int_{\R^{2n}} d\bs{\Delta} c\lr{\bs{\alpha},  \bs{\alpha}-\bs{\Delta}}\nu_{\sigma^2}\lr{\bs{\Delta}} \\
    &=\frac{1}{2}\lrq{1+\frac{\det\lr{\CL^{\perp}}}{2^{2n}}}.
    \nonumber
\end{align}

If the GKP code is chosen so to encode $k$ qudits each of dimension $d$ in $n$ modes, this computation estimates that the logical information accessible from the samples can be understood to have undergone a logical depolarizing channel with depolarizing probability
\begin{equation}
    p\leq \frac{1}{2}\lrq{1- 2^{-n \lr{2 + \frac{k}{n}\log_2\lr{d}} } }.
\end{equation}
For a fixed number $n$ of modes this estimate is reassuring: it tells us that the parameters can be chosen so that the effective logical error probability is bounded away from $\frac{1}{2}$. Asymptotically for large $n$, however, we are only promised an exponentially small amount of logical information content to survive the procedure. 

\subsubsection{A quasi-logical estimator}

The pointer states output by photon parity measurements are generically relatively unwieldy states and we expect it to be difficult to use them in a similar manner as e.g. the heterodyne pointers discussed earlier. Here it is more interesting to use samples from the Wigner function of an arbitrary state in the following way. 
Let $G\lr{\bs{x}}=W_G\lr{\bs{x}}$ be the Wigner function of an arbitrary 
\footnote{Hence, this is not necessarily finite dimensional.} trace-class observable on a system of $n$
quantum harmonic oscillators
and let $\CL\subseteq \CL^{\perp}$ describe a GKP code with even symplectic Gram matrix $A$. We define the estimator
\begin{equation}
    \widetilde{G}\lr{\bs{x}}  \coloneqq  W_{\rho}\lr{\bs{x}} G\lr{\bs{x}} \label{eq:estimator}
\end{equation}
such that samples from the Wigner function of an input state according to the distribution $p_{\sigma}\lr{\bs{\gamma};\CL^{\perp}}$can be combined with the observable Wigner function $G\lr{\bs{x}}$ at the same points to produce the expectation value
\begin{align}
    \overline{G}^{\sigma}_{\CL^{\perp}} &=  \int d\bs{x}\, p_{\sigma}\lr{\bs{x};\CL^{\perp}}  W_{\rho}\lr{\bs{x}} G\lr{\bs{x}}  \nonumber\\
    &\xrightarrow{\sigma \rightarrow 0}\sum_{\bs{\xi}^{\perp} \in \CL^{\perp}} W_{\rho}\lr{\bs{\xi}^{\perp}} G\lr{\bs{\xi}^{\perp}} \eqqcolon \overline{G}_{\CL^{\perp}}.\label{eq:sum_lat}
\end{align}
If either the Wigner function of the input state or the observable were solely supported on the lattice $\CL^{\perp}$, we see that in this limit $\sigma \to 0$ the expectation value 
\begin{equation}
    \Tr\lrq{\rho G}=\int_{\R^{2n}} d\bs{x}\, W_{\rho}\lr{\bs{x}} G\lr{\bs{x}}
\end{equation}
would be exactly reproduced. We discuss an interpretation of eq.~\eqref{eq:sum_lat} in the next subsection and will later see how the expression can be of use. 
First, we examine how well the lattice sum is approximated for small $0<\sigma \ll 1$. In order to bound the effective convergence of $p_{\sigma}\lr{\bs{x};\, \CL^{\perp}}\to \Sha_{\CL^{\perp}}\lr{\bs{x}}=\sum_{\bs{\xi}^{\perp}\in \CL^{\perp}} \delta\lr{\bs{x}-\bs{\xi}^{\perp}}$ under the integral, we need to make soft assumptions on the state 
and observable.

\begin{lem}[Random 
lattice point sampling]\label{lem:rdn_lat_points}
    Let $\rho, G$ be operators such that the product of their Wigner functions $ \widetilde{G}\lr{\bs{x}}  =  W_{\rho}\lr{\bs{x}} G\lr{\bs{x}}$ is well defined, $\Tr\lrq{G^2}<\infty$ is finite and further assume that 
    $W_{\rho}\lr{\bs{x}}, G\lr{\bs{x}}$ 
    are Lipschitz-continuous, with $\|\nabla W_{\rho}\lr{\bs{x}}\| \leq l_{\rho}$ and $\|\nabla G\lr{\bs{x}}\| \leq l_{G}$. Set $\tilde{l}=l_{\rho}+l_G$.
    Let $\sigma \ll \lambda_1\lr{\CL^{\perp}}$ be a small parameter. It holds that
    \begin{align}
    \overline{G}^{\sigma}_{\CL^{\perp}} &=  \int_{\R^{2n}} d\bs{x}\, p_{\sigma}\lr{\bs{x};\CL^{\perp}}  \widetilde{G}\lr{\bs{x}}  \nonumber \\
    &=\sum_{\bs{\xi}^{\perp} \in \CL^{\perp}} \widetilde{G}\lr{\bs{\xi}^{\perp}} +\epsilon\lr{\sigma} \\
    &= \overline{G}_{\CL^{\perp}}+\epsilon\lr{\sigma} \nonumber
    \end{align}
    with 
    \begin{equation}
        |\epsilon\lr{\sigma}| \leq \sqrt{2}\sigma \frac{\sqrt{2} \Gamma\lr{n+\frac{1}{2}}}{  \Gamma\lr{n}} \tilde{l}.
    \end{equation}
  Furthermore, it holds that
    \begin{align}
\overline{G}^{\sigma, 2}_{\CL^{\perp}} &=  \int d\bs{x}\, p_{\sigma}\lr{\bs{x};\CL^{\perp}}  \widetilde{G}^2\lr{\bs{x}}  \nonumber \\
    &\leq  \lr{\frac{2}{\pi\sigma^2}}^n\Tr\lrq{G^2}.
    \end{align}
\end{lem}
\endproof

Note that for large $n$, we have that $\Gamma\lr{n+\frac{1}{2}}/\Gamma\lr{n}= \sqrt{n}+O\lr{n^{-1/2}}$ while $\lambda_1\lr{\CL^{\perp}}\propto \sqrt{n}$ is the maximally achievable shortest vector scaling for any family of lattices $\CL^{\perp}=\CL^{\perp}_n$. Picking $\sigma=o\lr{1/\sqrt{n}}$ essentially allows to reduce the error $\epsilon\lr{\sigma}$ error to a negligible amount.
Again here we observe a non-trivial scaling with the system size, contrasting the bounds obtained in ref.~\cite{Huang_2020}. Note that the above statement concerns the variance of the estimator, which in ref.~\cite{Huang_2020} is upper bounded by defining the \textit{shadow norm}. The shadow norm is generically upper bounded using the rank of the observable, which, for finite dimensional systems, scales at most exponentially in the system size. Here the behavior is qualitatively different as $\Tr\lrq{G^2}$ captures the variance of the observable in phase space, for a potentially infinite-rank observable. The phase space integral that evaluates the trace may be split into separate integrals over each of the $n$ subspaces that represent the individual modes, for which the values are each suppressed by the factor $2/\pi\sigma^2$.

\subsubsection{Interpreting $\overline{G}_{\CL^{\perp}}$ \& structure of general code states } \label{sec:interpretation}

We wish to find an operational interpretation of the estimator $\overline{G}_{\CL^{\perp}}$ in the limit $\sigma \to 0$, which can be understood as the expectation value $\overline{G}_{\CL^{\perp}}=\Tr\lrq{\rho|_{\CL^{\perp}}G}$ of a state $\rho|_{\CL^{\perp}}$ whose Wigner function has been projected to be exclusively supported on the lattice $\CL^{\perp}$. A naive expectation would be that the code space projection $\rho\mapsto \Pi_{\CL}\rho\Pi_{\CL}$ (or performing optimal quantum error correction)  could perform such a restriction. In appendix~\ref{app:Wigner} we show that this is indeed not the case and that the Wigner function of a GKP code state is also supported on points in $\CL^{\perp}/2$ outside of $\CL^{\perp}$. And even if those points were not present, the projection $\rho\mapsto \Pi_{\CL}\rho\Pi_{\CL}$ does not clearly correspond to a restriction of the Wigner function of $\rho$ onto the desired support.

Here we defined the code space projector of a GKP code associated to lattice $\CL\subset \R^{2n}$, which is generally given by \footnote{Note that, formally, $\Pi_{\CL}$ is not a physically well-defined object and needs to be replaced by appropriate physically regularized versions. It describes the correct support of the code space and satisfies ${\rm supp}\lr{\Pi_{\CL}^2}\subseteq {\rm supp}\lr{\Pi_{\CL}}$ in the phase space description of operators but, due to the infinite size of the lattice, is not normalized to $\Pi_{\CL}^2=\Pi_{\CL}$. In a more careful treatment one would replace $\Pi_{\CL}$ by a regularized version, e.g. by introducing a probability measure over $\CL$, and would consider the present treatment as limiting case. To ease the presentation, and since $\Pi_{\CL}$ behaves like a projector on physical states in phase space, we ignore this artifact.}
\begin{equation}
    \Pi_{\CL, \phi} = \sum_{\bs{\xi}\in \CL} e^{i\phi \lr{\bs{\xi}}} D\lr{\bs{\xi}}.
\end{equation}
The function $\phi$ specifies the phase associated with a particular choice of \textit{pivot basis}~\cite{Conrad_2022, burchards2024fiberbundlefaulttolerance} or \textit{gauge}~\cite{Royer_2022}. 
We omit indicating $\phi$ since it will not be relevant for the present discussion and simply write $\Pi_{\CL}$. In particular, for scaled GKP codes with even $d$ it holds that $\phi=0$, which we now also assume to be the case. 
Earlier we have already encountered the Weyl- or phase-space-point operators $\Omega\lr{\bs{x}}=D\lr{\bs{x}}\hat{\pi}D^{\dagger}\lr{\bs{x}}$. These operators implement a degree $\pi$ rotation of a state around the point $\bs{x}$, such that $\Omega^2\lr{\bs{x}}=I$, are complete and orthogonal as
\begin{equation}
    \int_{\R^{2n}} d\bs{x}\, \Omega\lr{\bs{x}}=2^{-n} I, \quad \Tr\lrq{\Omega\lr{\bs{x}}\Omega\lr{\bs{y}}}=2^{-2n}\delta\lr{\bs{x}-\bs{y}}.
\end{equation}
Similar to their expansion in displacement operators, this allows to express any Hermitian trace class operator as
\begin{equation}
    O=\int_{\R^{2n}}d\bs{x}\, 2^{2n}\Tr\lrq{\Omega\lr{\bs{x}} O} \Omega\lr{\bs{x}} =  \int_{\R^{2n}}d\bs{x}\, 2^{n}W_O\lr{\bs{x}} \Omega\lr{\bs{x}},
\end{equation}
where we used that $\Tr\lrq{\Omega\lr{\bs{x}} O}=2^{-n}W_O\lr{\bs{x}}$. They further satisfy 
\begin{equation}
    D\lr{\bs{y}}\Omega\lr{\bs{x}}D\lr{\bs{y}}=e^{-i4\pi \bs{y}^TJ\bs{x}}\Omega\lr{\bs{x}},
\end{equation}
such that we observe that 
the operation
\begin{equation}
    \Gamma\lr{O} = \int d\mu^{\Gamma}\lr{\bs{y}}\,  D\lr{\bs{y}}OD\lr{\bs{y}} \label{eq:Gamma}
\end{equation}
can be understood as a map
\begin{equation}
    W_O\lr{\bs{x}} \mapsto W_O\lr{\bs{x}}\lrq{\int d\mu^{\Gamma}\lr{\bs{y}} e^{-i4\pi\bs{y}^TJ\bs{x}}}.\label{eq:Gamma_W}
\end{equation}
This form of modifiying the Wigner function of the observable is similar to the modification of the characteristic function under a displacement twirl encountered in sec.~\ref{sec:twirl}. A striking difference is an additional factor of $2$ in the exponent, which has the effect that a projective factor in eq.~\eqref{eq:Gamma_W} only uniformly projects onto $\CL^{\perp}$ if we choose
\begin{equation}
    d\mu^{\Gamma}\lr{\bs{y}}=  d\bs{y}\Sha_{\CL/2}\lr{\bs{x}}.\label{eq:mu_Gamma}
\end{equation}
with the Dirac comb defined for lattice $L$ as
\begin{equation}
    \Sha_{L}\lr{\bs{x}}=\sum_{\bs{v}\in L}\delta\lr{\bs{x}-\bs{v}}.
\end{equation}
Combining this, the operation in eq.~\eqref{eq:Gamma} still does not generally correspond to a displacement twirl obtained from by randomly displacing an input state by vectors in $\CL/2$. This, however, will be the case if we further specify the following.
\begin{enumerate}
\item $\CL$ is such that $\forall \bs{\xi}, \bs{\xi}' \in \CL: \, \bs{\xi}^TJ\bs{\xi}' \in 4\Z$. This is such that two displacements by $\bs{z}, \bs{z}'\in \CL/2$ combine without additional phase $D\lr{\bs{z}}D\lr{\bs{z}'}=D\lr{\bs{z}+\bs{z}'}$.
\item[2.] The output state is evaluated on the code space of $\CL$, e.g., by performing syndrome measurements and post-selecting on zero syndrome on the input state or by projecting $G\mapsto \Pi_{\CL}G\Pi_{\CL}$ onto the logical observable associated to $G$. Together, this has the consequence that we can replace the rightmost displacement via $\bs{y}\in \CL/2$ in eq.~\eqref{eq:Gamma} by $D\lr{\bs{y}}\Pi_{\CL}=D\lr{\bs{y}}^{\dagger}\Pi_{\CL}$, so that the map is understood ad random displacement channel.
\end{enumerate}
Restricted to the code space of $\CL$, the infinite twirl specified in eq.~\eqref{eq:mu_Gamma} can be replaced by a uniform measure over the finite set $\frac{1}{2}\CL / \CL$. Observe that for a scaled GKP code with $\CL=d\CL^{\perp}$ we have in the case $d=2$ precisely that a twirl over $\frac{1}{2}\CL / \CL$ is logically maximally depolarizing. In consequence, the state is projected onto a contribution proportional to $\Pi_{\CL}$ and loses all information except for the proportionality constant. We hence focus on scaled GKP codes with $d\in 4\Z$, which also automatically satisfied point $1.$ above. In general, the code space projected state is specified by $|\CL^{\perp}/\CL|=d^{2n}$ coefficients. The additional twirl over displacements in $\frac{1}{2}\CL / \CL$ will only preserve characteristic function contributions in $\lr{\frac{1}{2}\CL}^{\perp}=2\CL^{\perp}\subset \CL^{\perp}$, which amounts to $|2\CL^{\perp}/\CL|=|\CL^{\perp}/2\CL^{\perp}|^{-1}|\CL^{\perp}/\CL|=(d/2)^{2n}$ coefficients specifying \textit{logical} information of the GKP code associated to $\CL/2$. 
We summarize: for scaled GKP codes with $d\in 4\mathbb{N}_{>1}$, the limit $\sigma \to 0$ of the estimator $\overline{G}_{\CL^{\perp}}^{\sigma}$ in eq.~\eqref{eq:sum_lat} encodes the logical expectation values of $G$ relative to $\rho$ for a GKP code specified by lattice $\CL/2$.

\section{Chasing shadows}\label{sec:chasing_shadows}

\begin{figure}
    \centering
    \includegraphics[width=0.7\linewidth]{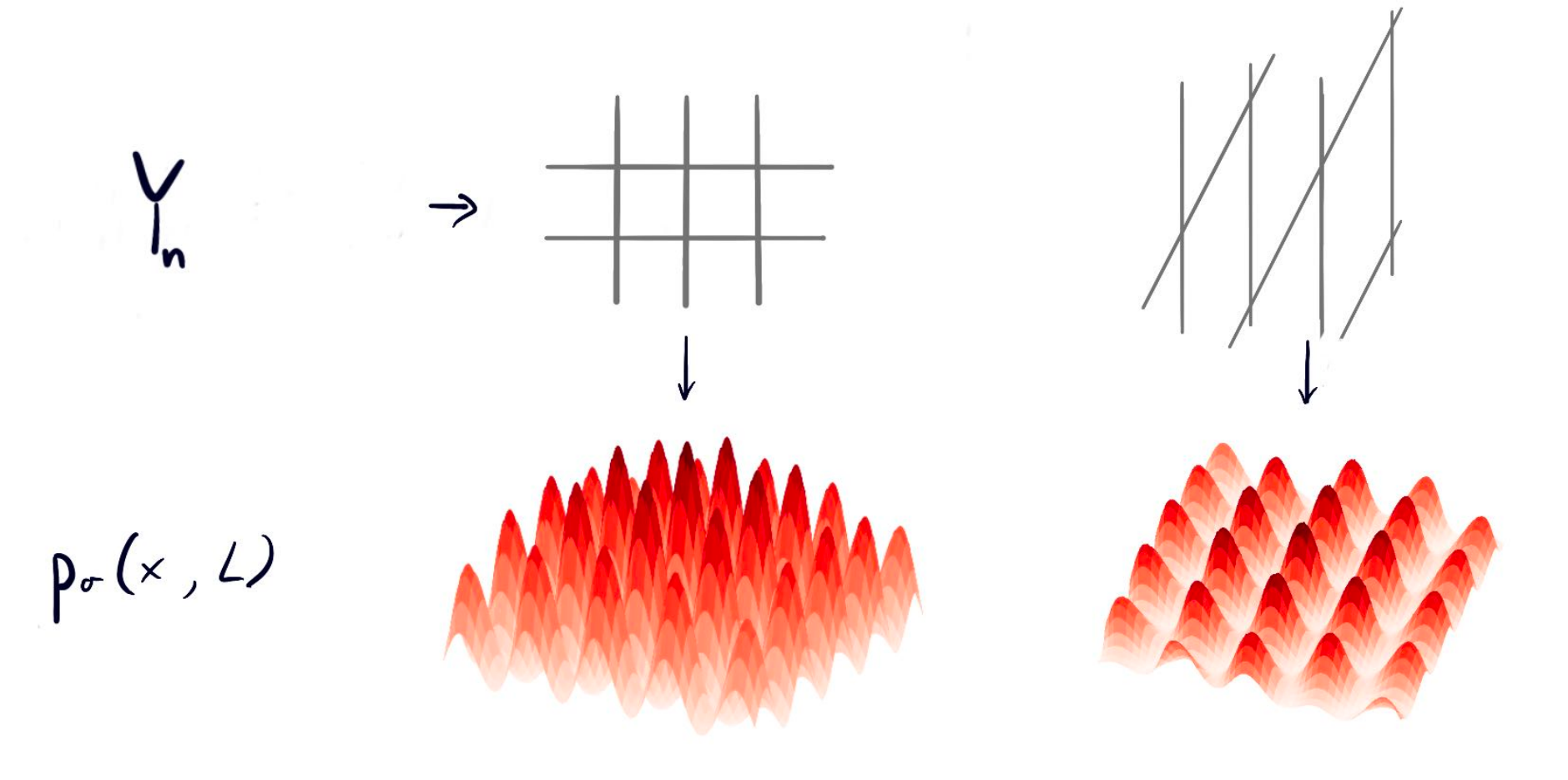}
    \caption{Phase-space distribution $p_{\sigma}(\cdot, L)$ for the GKP Wigner shadow protocol, which are determined by a randomly chosen lattice $L$.}
    \label{fig:lat_distributions}
\end{figure}

In this section, we investigate how the estimator derived in the previous section together with a random choice of lattice allows to construct a shadow tomography protocol for the Hilbert space of a $2n$-dimentional continuous variable system. Crucially, by averaging over lattices, the reference to any particular GKP code is removed. The following  presents a randomized scheme to select phase space points on which the Wigner function of an arbitrary input state are evaluated so to estimate any observable without restriction to any particular GKP code space (see fig.~\ref{fig:lat_distributions} for an illustration of such randomly selected distributions).  
We continue to focus on scaled GKP codes obtained from rescaling a symplectically self-dual lattices $L=L^{\perp}\subseteq \R^{2n}$. Such a lattice is spanned by the rows of a symplectic generator matrix $M\in \Sp_{2n}\lr{\R}$ and the space of all possible symplectic lattices in dimension $2n$ is simply parametrized by the group of symplectic matrices \textit{up to basis transformations} $Y_{n}=\Sp_{2n}\lr{\Z}\backslash \Sp_{2n}\lr{\R}$, where here the ``$\backslash$'' operator is meant to denote a left-modulo operation since basis transformations are implemented by left-multiplication on a given generator matrix.

\subsection{Moments of symplectic lattices}

Following work by Siegel and Rogers, it has been realized by Buser and Sarnak in ref.~\cite{Sarnak1994} and by Kelmer and Yu in ref.~\cite{Kelmer_2019} that the space of symplectic lattices possesses a Haar measure $\mu\lr{L}$, relative to which functions of the lattice can be integrated. Define the Siegel transform of a sufficiently fast decaying function $f:\R^{2n}\to \R$ by

\begin{equation}
    F_f\lr{L}=\sum_{\bs{v}\in L_{\rm pr}} f\lr{\bs{v}},
\end{equation}
where $L_{\rm pr}$ denotes the set of primitive vectors of $L$, that is the set of minimal vectors that cannot be obtained by another via integer multiplication. This (infinite) set is such that every point in the lattice $L=\N \otimes L_{\rm pr}$ is uniquely reproduced a positive integer multiple of a point in $L_{\rm pr}$.
We also define

\begin{equation}
    \widetilde{F}_f\lr{L}=\sum_{\bs{v}\in L-\lrc{0}} f\lr{\bs{v}}=\sum_{k\in \N} \sum_{\bs{v}\in L_{\rm pr}} f\lr{k \bs{v}}.
\end{equation}
The symplectic version of Siegel's mean value theorem, as derived in ref.\ \cite{Sarnak1994}, can thus be formulated as
\begin{equation}
    \int_{Y_n} d\mu\lr{L} F_f\lr{L} = \frac{1}{\zeta\lr{2n}}\int_{\R^{2n}}d\bs{x}\, f\lr{\bs{x}},
\end{equation}
with $\zeta\lr{z}=\sum_{k\in \N }k^{-z}$ 
being 
the Riemann zeta function. Applying this formula once allows to derive its perhaps more standard variant
\begin{align}
    \int_{Y_n} d\mu\lr{L} \widetilde{F}_f\lr{L}&= \frac{1}{\zeta\lr{2n}}\sum_{k\in \N} \int_{\R^{2n}} d\bs{x} f\lr{k\bs{x}} \\ &=  \int_{\R^{2n}} d\bs{x} f\lr{\bs{x}}.
    \nonumber
\end{align}
Analogous to the inner product over $L^2\lr{\R^{2n}}$, \begin{equation}
    \braket{f, g}=\int_{\R^{2n}} d\bs{x}f\lr{\bs{x}}^*g\lr{\bs{x}},
\end{equation}
one can define an inner product over $Y_n$ as
\begin{equation}
    \braket{F, G}_{Y_n}=\int_{Y_n} d\mu\lr{L}F\lr{L}^*G\lr{L}.
\end{equation}
In this notation, the mean value formula can be simply expressed as 
\begin{equation}
    \braket{1, F_f}_{Y_n}=\frac{1}{\zeta\lr{2n}}\braket{1, f}.
\end{equation}
In ref.~\cite{Kelmer_2019}, Kelmer and Yu have derived a formula for second moments over $Y_n$, which for two even and compactly supported functions $f, g$ can be compactly written as 

\begin{equation}
    \braket{F_f, F_g}_{Y_n}=\frac{\braket{f, 1}\braket{1, g}}{\zeta\lr{2n}^2} + \frac{2}{\zeta\lr{2n}}\lr{\braket{f, g}+\braket{\iota \lr{f}, g}},
\end{equation}
where $\iota$ is an isometry such that $\|\iota \lr{f}\|=\|f\|=\sqrt{\braket{f, f}}$.
Note that, since $L_{\rm pr}$ and $L$ are even sets, any non-even function $f$  on the LHS of this formula can simply be replaced by their even projection $\frac{f\lr{\bs{x}}+ f\lr{-\bs{x}}}{2}$ (similar for $g$): the assumption of evenness of the functions is without loss of generality.
Using this expression, we show the following statement in the appendix.

\begin{lem}[Scalar product bound]\label{lem:scalar}
    Let $f, g$ be two even compactly supported functions, it holds that
    \begin{equation} \left|\braket{\widetilde{F}_f, \widetilde{F}_g}_{Y_n}\right| \leq |\braket{f, 1}\braket{1, g}|+\frac{4\zeta\lr{n}^2}{\zeta\lr{2n}}\|f\|\|g\|.
    \end{equation}
\end{lem}

\subsection{Estimating CV observables with random lattices}

The mean value theorem and Lemma~\ref{lem:scalar} are the key 
tools to analyse the behaviour of the estimator $\widetilde{G}$ when in addition to the points $\bs{x}\sim p_{\sigma}\lr{\bs{x}, L}$ being randomly sampled from a lattice-Gaussian-like distribution also the lattice $L=L^{\perp}$ is chosen according to a uniformly random distribution $\mu\lr{L}$ over the space of lattices $L=L^{\perp}$.
Under this procedure, the estimator takes the expectation value 
\begin{align}
    \int_{Y_n} d\mu\lr{L}\int_{\R^{2n}} d\bs{x}\, p_{\sigma}\lr{\bs{x};\, L} \widetilde{G}\lr{\bs{x}}&= \int_{Y_n} d\mu\lr{L}\,\overline{G}_L \nonumber\\ 
    &= \widetilde{G}\lr{0} + \int_{\R^{2n}} d\bs{x}\, \widetilde{G}\lr{\bs{x}}, \nonumber\\
    &= \widetilde{G}\lr{0} + \Tr\lrq{\rho G}\nonumber \\
    &\eqqcolon \overline{G}
\end{align}
which is attained up to a small error $\epsilon\lr{\sigma}$ as in Lemma~\ref{lem:rdn_lat_points}. 
Note, that $\widetilde{G}\lr{0}=\Tr\lrq{\hat{\pi}\rho}\Tr\lrq{\hat{\pi}G}$ is simply the total photon parity of the state and observable together and can be offset by an independent estimation of the photon parity of the unknown input state.

We consider the following two scenarios. In the first, we simply sample a point $\bs{x}_{i, k}$ according to the distribution $p_{\sigma}\lr{\bs{x}, L_k}$ for each randomly chosen lattice $L_k$. In this case a second moment bound is obtained as follows.

\begin{lem}[Second moment: random lattice sampling]\label{lem:random_lat_1}
    Let $\tilde{G}$ be an estimator as in Lemma~\ref{lem:rdn_lat_points} for an observable on a $n$-mode continuous variable quantum system,  symplectic  lattice $L=L^{\perp}\subset\R^{2n}$ and $d\in \mathbb{N}$ a natural number. Assume that the observable $G$ is such that 
    \begin{equation}
        \|G\|^2_2\coloneqq \int d\bs{x}\, |G\lr{\bs{x}}|^2<\infty 
    \end{equation}
    is finite. 
It holds that
\begin{equation}
    \int_{Y_n}d\mu\lr{L}\int d\bs{x}\, p_{\sigma}\lr{\bs{x}; d^{-1/2}L} \widetilde{G}\lr{\bs{x}}^2 \leq \lr{\frac{d}{2\pi}}^{n}\lrc{1+c_{\sigma, d}^n}\|G\|_2^2 \coloneqq V_1\lr{G, n, d, \sigma}, 
\end{equation}
with $c_{\sigma, d}=\frac{2\pi d}{\sigma^2+\sigma^{-2}}\in [0, d\pi]$
\end{lem}
We have $c_{\sigma, d}\leq 1$ for $\sigma^2\leq \pi d\pm \sqrt{\pi^2d^2-1}$. 
The second scenario that we consider is such that the ``inner'' average is computed with high precision so that the only contributions to the variance arise from the choice of random lattices. Using similar techniques and Lemma~\ref{lem:scalar}, in appendix~\ref{app:lem4} we show that estimation of $\overline{G}_L^{\sigma}$ in this setting has bounded variance as follows.
\begin{lem}[Second moment: random lattice sampling for exact inner means]\label{lem:random_lat_inner}
Let $\tilde{G}$ be an estimator as in Lemma~\ref{lem:rdn_lat_points} for an observable on a $n$-mode continuous variable quantum system with $n>1$, $L=L^{\perp} \subset \R^{2n}$ denote a symplectic lattice and let $d\in \mathbb{N}$ be a natural number.
 Assume that the observable $G$ is such that 
    \begin{equation}
        \|G\|_1 \coloneqq \int d\bs{x}\, |G\lr{\bs{x}}|<\infty 
    \end{equation}
    is finite. 
   Let  
   \begin{equation}
       \overline{G}_{\CL^{\perp}}^{\sigma}= \int d\bs{x}\, p_{\sigma}\lr{\bs{x};\, \CL^{\perp}} \widetilde{G}\lr{\bs{x}}
   \end{equation}
   be the estimator from Lemma~\ref{lem:rdn_lat_points} with $\CL^{\perp}=d^{-1/2}L$ and
    let $\sigma>0$. It holds that
   \begin{equation}
        \left|\int_{Y_n} d\mu\lr{L}\lr{\overline{G}_L^{\sigma}}^2\right|\leq  \lr{\frac{d}{4\pi}}^{2n}\lrq{1+\lr{2+2^{2-n}\frac{\zeta(n)^2}{\zeta(2n)}}c_{\sigma, d}^n+c_{\sigma, d}^{2n}}\|G\|_1^2
    \end{equation}
    with $c_{\sigma, d}=\frac{2\pi d}{\sigma^2+\sigma^{-2}}\in [0, d\pi]$.
\end{lem}

This second moment bound naturally implies a variance bound for the expectation value $\overline{G}$ in the scenario where the inner average is taken with high precision and the only fluctuations around the mean essentially arise from the choice of random lattice. We compare this behavior with the scaling in lem.~\ref{lem:random_lat_1} for small $\sigma < \pi d-\sqrt{\pi^2d^2-1}$ with $n$. First note that the factor $\zeta(n)^2/\zeta(2n)\xrightarrow{n\to \infty} 1$ converges rapidly to a constant. We can hence approximate to low order in $c_{\sigma, d}$
\begin{equation}
   \lr{4\pi}^{-2n}\lrq{1+\lr{2+2^{2-n}\frac{\zeta(n)^2}{\zeta(2n)}}c_{\sigma, d}^n+c_{\sigma, d}^{2n}}\|G\|_1^2
   \approx \frac{1+2c_{\sigma, d}^n}{(4\pi)^{2n}} \|G\|_1^2.\label{eq:rand_lat_inner}
\end{equation}
The factor of $(4\pi)^{-2n}$ suggests an exponentially faster decay of the prefactor as compared to the bound in lem.~\ref{lem:random_lat_1}. On the other hand, for large $\sigma$ such that $c_{\sigma, d}>1$, the dominant contribution of eq.~\eqref{eq:rand_lat_inner} scales as
\begin{equation}
    \lr{\frac{c_{\sigma, d}}{4\pi}}^{2n}\|G\|_1^2 \leq \lr{\frac{d}{4}}^{2n}\|G\|_1^2,
\end{equation}
where we finally used that it generally holds that $c_{\sigma, d}\leq \pi d$. This suggests that the estimator variance of the second scenario generally remains well-behaved and favorable relative to the previous setting for $d\leq 4$, but $\sigma$ needs to be tuned sufficiently small in all other cases. A further relevant difference is in the observable norm appearing in these bounds. While the $2$-norm, appearing in lem.~\ref{lem:random_lat_1} captures the phase-space variance of the observable $G$, the bound in lem.~\ref{lem:random_lat_inner} depends on its one-norm, which quantifies the \textit{Wigner-negativity} of the observable $G$ when its Wigner function is normalized \cite{Hudson19742}.

\subsection{Protocols}
We outline two shadow estimation protocols for arbitrary states and appropriate observables, also described in figs.~\ref{prot:CV_shadow} and \ref{prot:CV_shadow2}. Both protocols build on sampling points $\bs{x} \xleftarrow{p_{\sigma}\lr{\cdot, \CL^{\perp}}} \R^{2n}$ according to the probability density $p_{\sigma}\lr{\cdot, \CL^{\perp}}$, relative to a random lattice $L$, corresponding to a scaled GKP code with $\CL=\sqrt{d}L$ and $\CL^{\perp}=L/\sqrt{d}$, on which the Wigner function of the state is to be evaluated, e.g., by performing displaced photon parity measurements. To achieve a high precision in this step, an extra constant overhead will be required, which we however suppress in our notation. 
This is the \textit{POVM twirl} in this shadow tomography setting and the Wigner function values evaluated over these point samples form the classical shadow of this protocol. In a post-processing step, they can be combined with selected observables (i.e, their Wigner function) to estimate its expectation.

\subsubsection{Protocol 1} Protocol~\ref{protocol1} proceeds by sampling a lattice form the uniform measure $\mu(L)$ over $Y_n$, the space of all symplectic lattices. For each sampled lattice, the probability density $p_{\sigma}(\bs{x}, \CL^{\perp})$ is used to sample a point $\bs{x} \in \R^{2n}$, which, by construction, is close (with variance $\sigma^2$) to the lattice $\CL^{\perp}$. The experiment evaluates the Wigner function at this point and stores its value $\lr{\bs{x}, W_{\rho}\lr{\bs{x}}}$ together with the sampled point as one data point of the classical shadow. For an appropriate observable $G$, later estimation then proceeds by using median of means estimation on the sample set $\lrc{\widetilde{G}\lr{\bs{x}_i}=W_{\rho}\lr{\bs{x}_i}G\lr{\bs{x}_i}}_{i=1}^N$.
The estimator variance here is captured by lemma~\ref{lem:random_lat_1}, which provides an upper bound on its second moment (and hence also its variance). In this protocol the random lattice simply serves as an auxiliary object to sample points in a well-distributed manner from $\R^{2n}$.
\begin{figure*}
    \begin{mybox}
        \begin{prot}\label{protocol1}
            \begin{enumerate}
                \item[]
                \item Sample $N=CKB$ points  $\bs{x}_i \xleftarrow{p_{\sigma}\lr{\bs{x}, \CL^{\perp}_k}} \R^{2n}\supset \CL^{\perp}_k=d^{-1/2}L_k \xleftarrow{\mu\lr{L}}Y_n$, where
                \begin{align*}
                    K&=2\log\lr{2M/\delta}\\
                    B&=\frac{1+c_{\sigma, d}^n}{(2\pi)^n} \max_{m}\|G_m\|_2^2/\tilde{\epsilon}^2
                \end{align*}
                and $C$ is a constant.
                \item Evaluate the Wigner function of an unknown input states at these points $\lrc{\bs{x}_i}_{i=1}^N$ and combine the output with the Wigner function of the observable $G\lr{\bs{x}_i}$ at these points.
                \item For each point, return the estimate  $\widetilde{G}\lr{\bs{x}_i}=W_{\rho}\lr{\bs{x}_i}G\lr{\bs{x}_i}$.
             \end{enumerate}
            \end{prot}
            \caption{Random CV shadow protocol.}\label{prot:CV_shadow}
    \end{mybox}
\end{figure*}

\subsubsection{Protocol 2} Protocol~\ref{protocol2} proceeds similar to protocol~\ref{protocol1}, but rather than sampling only one point $\bs{x} \xleftarrow{p_{\sigma}\lr{\cdot, \CL^{\perp}}} \R^{2n}$ for every sampled lattice $L=\sqrt{d}\CL^{\perp}$, we try to obtain a good estimate for
   \begin{equation}
       \overline{G}_{\CL^{\perp}}^{\sigma}= \int_{\R^{2n}} d\bs{x}\, p_{\sigma}\lr{\bs{x};\, \CL^{\perp}} \widetilde{G}\lr{\bs{x}}
   \end{equation}
for each sampled lattice. According to lemma~\ref{lem:rdn_lat_points}, this estimation step is governed by the variance upper bound (again using the second moment as proxy) 
\begin{equation}
    V_{1}(G; n, \sigma) = \lr{\frac{d}{2\pi}}^n\lrc{1+c_{\sigma, d}^n} \|G\|_2^2, \; c_{\sigma, d}=\frac{2\pi d}{\sigma^2+\sigma^{-2}}\in \lrq{0,\pi d}. 
\end{equation}
For each individual sampled lattice, $O\lr{\lr{2/\pi\sigma^2}^n \Tr\lrq{G^2}}$ samples suffice in suppressing the additive error on $\overline{G}_{\CL^{\perp}}^{\sigma}$ to constant average variance, independent of the chosen lattice. 
The total estimation error from a sequential protocol, 1. sampling a lattice $L\xleftarrow{\mu} Y_n$ and 2. sampling a point $\bs{x}\xleftarrow{p_\sigma(\cdot, \CL^{\perp})} \R^{2n}$ is additive. 

Together with the bound in lemma~\ref{lem:random_lat_inner} we hence obtain the following.
\begin{cor}[Variance upper bound of protocol~\ref{protocol2}]
Let $\tilde{G}$ be an estimator as in Lemma~\ref{lem:rdn_lat_points} for an observable on a $n$-mode continuous variable quantum system where $n>1$ and a GKP code with $\CL=\sqrt{d}L$ given by symplectic lattice $L=L^{\perp}\subset\R^{2n}$.
     Assume that the corresponding observable $G$ is such that 
    \begin{equation}
        \|G\|_1 \coloneqq \int_{\R^{2n}} d\bs{x}\, |G\lr{\bs{x}}|<\infty 
    \end{equation}
    is finite. 
   Let  
   \begin{equation}
       \overline{G}_{\CL^{\perp}}^{\sigma}= \int_{\R^{2n}} d\bs{x}\, p_{\sigma}\lr{\bs{x};\, \CL^{\perp}} \widetilde{G}\lr{\bs{x}}
   \end{equation}
   be the estimator from Lemma~\ref{lem:rdn_lat_points} and
    let $\sigma>0$. It holds that the estimator of protocol~\ref{protocol2} has variance upper bounded by

    \begin{equation}
        \mathrm{Var}\lr{\overline{G}_{\rm 2}}\leq \lr{\frac{2}{\pi\sigma^2}}^n\Tr\lrq{G^2}+\lr{\frac{d}{4\pi}}^n\lrq{1+\lr{2+2^{2-n}\frac{\zeta(n)^2}{\zeta(2n)}}c_{\sigma, d}^n+c_{\sigma, d}^{2n}}\|G\|_1^2\coloneqq V_2\lr{G; n, d,  \sigma}.
    \end{equation}
    
\end{cor}

It is relevant to keep in mind that protocols~\ref{protocol1} and \ref{protocol2} estimate the expectation value of $\Tr\lrq{\rho G}$ only up to the photon parity offset $\widetilde{G}(0)$. Correcting for this offset, assuming it can be pre-estimated with high confidence, however has no effect on the variances of the estimators. We can combine the variance estimates with the shadow tomography protocol of Huang et al.~\cite{Huang_2020}, which, adapted to out setting, proceeds by a median of means strategy to quantify how the the data obtained from our protocols can be used to estimate a large number of observables. We state the conclusions here and provide the relevant proofs in the appendix. 

\begin{them}[Random lattice CV shadows]\label{them:Randomlatshadow}
     Let $\tilde{\epsilon}, \delta, \sigma> 0$ be small parameters, $d\in \mathbb{N}$ an integer, let $G_{m},\, m=1,\hdots, M$ be operators with finite $\|G_m\|_2^2$ and set $K, B$ as described in protocol~\ref{protocol1}.
    $N=CKB$, samples from the distribution of phase space points $\lrc{\bs{x}_i}_{i=1}^N$ sampled according to protocol~\ref{protocol1} approximate the expectation values 
    \begin{equation}
        \overline{G}_{m}=\Tr\lrq{\rho G_m},
        m=1,\hdots,M
    \end{equation}
     of an arbitrary state on an $n$-mode continuous variable quantum system 
     up to photon parity offsets $\tilde{G}_m\lr{0}$ and error $\tilde{\epsilon}$ 
     with probability at least $1-\delta$.
\end{them}

\begin{them}[Random lattice CV shadows with exact inner means]\label{them:Randomlatshadow2}
     Let $\tilde{\epsilon}, \delta, \sigma> 0$ be small parameters, $d\in \mathbb{N}$ an integer, let $G_{m},\, m=1,\hdots, M$ be operators with finite $\|G_m\|_2^2$ and set $K, B$ as described in protocol~\ref{protocol2}.
    $N=CC'KBN_P$, samples from the distribution of phase space points $\lrc{\bs{x}_i}_{i=1}^N$ sampled according to protocol~\ref{protocol2} approximate the expectation values 
    \begin{equation}
        \overline{G}_{m}=\Tr\lrq{\rho G_m},
        m=1,\hdots,M
    \end{equation}
     of an arbitrary state on an $n$-mode continuous variable quantum system 
     up to photon parity offsets $\tilde{G}_m\lr{0}$ and error $\tilde{\epsilon}$ 
     with probability at least $1-\delta$.
\end{them}

\begin{figure*}
    \begin{mybox}
        \begin{prot}\label{protocol2}
               \begin{enumerate}
                \item[]
                \item Sample $N_L= C KB$ lattices $\CL_k=\sqrt{d}L_k$, where $L_k \xleftarrow{\mu\lr{L}}Y_n$ is sampled uniformly and, where          \begin{align*}
                    K&=2\log\lr{2M/\delta},\\
                B&= \lr{4\pi}^{-2n}\lrq{1+\lr{2+2^{2-n}\frac{\zeta(n)^2}{\zeta(2n)}}c_{\sigma, d}^n+c_{\sigma, d}^{2n}}\max_m\|G_m\|_1^2 /\tilde{\epsilon}^2,
                \end{align*}
                with $C$ a constant.
                For each lattice sample 
                \begin{equation}
                    N_P=C'\lr{2/\pi\sigma^2}^n\max_m \Tr\lrq{G_m^2}
                \end{equation}
                points
                $\bs{x}_{kN_P+i} \xleftarrow{p_{\sigma}\lr{\bs{x}, \CL^{\perp}_k}} \R^{2n}, k=0, \dots, N_L-1,\, i=0, \dots, N_P-1$,
                where $C'$ is another constant. 
                
                \item Evaluate the Wigner function of 
                an unknown input states at these points $\lrc{\bs{x}_i}_{i=1}^{N_LN_P}$. 
                \item To estimate an observable, combine the output with the Wigner function of the observable $G\lr{\bs{x}_{i}}$ at these points.
                \item For each $k=0, \dots, N_L-1$, return the estimate  
                \begin{equation}
                    \widetilde{G}_{\CL^{\perp},k}^{\sigma} = N_P^{-1} \sum_{i=0}^{N_P-1} W_{\rho}\lr{\bs{x}_{kN_P+i}}G\lr{\bs{x}_{kN_P+i}}.
                \end{equation}
             \end{enumerate}
            \end{prot}
            \caption{Random CV shadow protocol.}\label{prot:CV_shadow2}
    \end{mybox}
\end{figure*}

These protocols describe particularly structured ways to sample phase space points $\bs{x}\in \R^{2n}$ to determine Wigner shadows. The significance of these protocols lies in the properties of the Haar measure over the space of symplectic lattices, which allow variances of the resulting estimation procedure to be well-quantified with minimal assumptions on the state and observables. A further feature is that the parameter $\sigma$ can serve as an interesting dial on this protocol. While, in the limit $\sigma \to \infty$, the distribution $p_{\sigma}$ simply becomes increasingly flat, which may also be a choice in an naive attempt to sample points from phase space under physical assumptions on state and/or observable. The opposite limit $\sigma \to 0$ is instructive in that the distribution $p_{\sigma}\lr{\cdot, \CL^{\perp}}$ tends to a narrow Gaussian distribution around the lattice points of $\CL^{\perp}$. Such distributions have significance in cryptography~\cite{Bernstein2008}, and more importantly, resemble the structure of GKP states~\cite{GKP}. This motivates to investigate how such resource state, or their induced ability to perform GKP error correction \cite{GKP, Glancy_2006} can serve as a tomographic resource. 

In sec.~\ref{sec:interpretation} we have discussed the interpretation of the estimator 
\begin{equation}
    \overline{G}_{\CL^{\perp}}^{\sigma}= \int_{\R^{2n}} d\bs{x}\, p_{\sigma}\lr{\bs{x};\, \CL^{\perp}} \widetilde{G}\lr{\bs{x}} \xrightarrow{\sigma \to 0} \sum_{\bs{\xi}^{\perp}\in \CL^{\perp}} W_{\rho}(\bs{\xi}^{\perp})G(\bs{\xi}^{\perp}),\label{eq:est_sigma_to_0}
\end{equation}
in the $\sigma \to 0$ limit as the logical expectation value of the observable $G$ under a GKP code specified by lattice $\CL/2$. To obtain this perspective we assumed $\CL$ to correspond to a scaled GKP code $\CL=d\CL^{\perp}$ with scaling parameter $d\in 4\mathbb{N}_{>1}$. 

Comparing to lemma.~\ref{lem:rdn_lat_points}, in this setting we observe that idealized GKP error correction allows to estimate the inner means $\overline{G}_{\CL^{\perp}}$ of protocol~\ref{protocol2} exactly, such that the ability to perform code space projections (or implementing optimal GKP quantum error correction) can be understood as a resource equivalent to additional sample complexity of asymptotic size $ \lim_{\sigma \to 0} O\lr{\exp\lr{(2/\pi\sigma^2)^n}}$, which is otherwise necessary to obtain an estimate for $\overline{G}_L^{\sigma}$ with perfect confidence. While we do not further explore this here, it would be interesting to investigate whether a physical protocol can be designed that estimates $G^{\sigma}_L$ for $\sigma >0$.

\section{Further related work}

It is a valid question to ask whether involved procedures as designed in this work really are necessary to design a shadow protocol for observables on a CV system. Many alternative strategies seem equally valid, in particular through alternative means to regularize the Hilbert space to an effectively finite-dimensional system. This is the approach that has been taken by Iosue et al. and Ghandari et al.\ in refs.~\cite{iosue2022continuousvariable, gandhari2022continuousvariable}. A related technical ingredient to draw random unitaries on a CV Hilbert space is also presented in ref.~\cite{Zhong2023information}. The interesting observation to be taken away from this section is that it is in fact possible to interpolate between \textit{logical} shadow tomography of a discrete quantum system embedded in a CV Hilbert space and \textit{physical} shadow tomography applied to the full CV Hilbert space. The bounds derived in this section explain exactly the cost of this interpolation. This property appears to be special to GKP codes and is implied by the fact that the group algebra of displacement operators defining the GKP stabilizer group together by their real powers form a complete operator basis for the CV Hilbert space and the existence of a Haar measure over the possible codes. These are properties not shared by other bosonic quantum error correcting codes that we know of where such a interpolation does not seem to be possible.
Note that in ref.~\cite{CVLearning} an information theoretic bound for learning CV quantum states 
has been derived, which we recall.

\begin{them}[Obstructions against general quantum state tomography \cite{CVLearning}]\label{them:CVlearning}
Let $\rho$ be an unknown quantum state on $n$ bosonic 
modes satisfying an energy
constraint $\Tr(\hat N\rho)
\leq n \overline{N}$ for some absolute constant $\overline{N}$. Then the number of copies of $\rho$ required to perform quantum state tomography with precision $\varepsilon$ in trace distance has to scale at least as 
$(\Theta (\overline{N}/\varepsilon))^{2n}$.
\end{them}

From the sample overhead scaling in Theorem~\ref{them:Randomlatshadow}, 
we clearly see that our protocol appears consistent with this bound, scaling both 
exponential in the system size as well as the lattice approximation parameter $\sigma^{-1}$. We can model an average energy constraint on an input state by assuming it to be in the support of a Gaussian regularizer $\hat{R}=e^{-\sigma\hat{N}}$, which constrains the typical support of any input state to be within a phase-space radius of 
\begin{equation}
r=\Omega\lr{\sqrt{n \sigma^{-1}}}. 
\end{equation}
In a rough estimate, we relate the typical support radius of such a state in phase space $r^2\propto n\overline{N}$ to its average photon number, which conversely leads us to the estimate that, given a state with average photon number constrained as above, sampling the Wigner function from a probability distribution with Gaussian envelope with variance $\sigma^{-1} \propto \overline{N}$ suffices to cover the essential support of the state. With this estimate the theorem above thus expects a scaling of the sample overhead with $\sigma^{-2n}\propto N_P$, stemming (for small $\sigma$) mainly from the ``inner'' estimation procedure. In contrast, when the inner estimator $\overline{G}_L^{\sigma}$ is obtained with high confidence, the average over different lattices only incurs an overhead growing polynomially in $c_{\sigma, d}\approx 2\pi d\sigma^2$, while remaining exponential in $n$ for sufficiently large choice of parameter $d$.

In ref.~\cite{conrad2025continuousvariabledesignsdesignbasedshadow} it was moreover shown that the space of GKP codes yields a rigged CV 2-design~\cite{iosue2022continuousvariable} and that this property also allows to derive CV shadow tomography protocols.  The key concept of interest presented here is that our scheme allows to interpolate between encoded logical information relative to a random code and physical tomographic data of the CV system and is used to construct an explicit randomized Wigner tomography protocol.

\section{Discussion and outlook}\label{sec:DiscussionOutlook}

In this article, we have developed the toolbox of shadow tomography for a continuous variable system relative to GKP codes, which themselves possess an intrinsic infinite structure and so that the infinitude of the CV phase space can be matched.  Our analysis and bounds on logical shadow tomography for GKP codes is particularly useful to design and test logical properties of experimental realizations of GKP code states such as pursued for GKP-\emph{measurement based quantum computations} 
(MBQC) \cite{Bourassa_2021_blue,  Menicucci_2014} or 
 realizations via superconducting architectures,  where a photon parity measurement is inexpensive \cite{Terhal_2020,  lachancequirion2023autonomous}. The twirling-centric perspective on shadow tomography highlighted in this manuscript complements perspectives obtained using the language of generalized POVMs and frame theory developed in refs.\  ~\cite{Innocenti, Acharya, Ngyuen, conrad2025continuousvariabledesignsdesignbasedshadow}. Our treatment exploits that this perspective neatly captures how encoded logical information presents itself in a physical POVM.

The logical channel twirl of a bosonic POVM introduced here is a powerful technique in its own right, as the space of bosonic POVMs is plentiful and one often does not have native access to a POVM tailored to the observable one desires to measure.  As we have discussed,  one may, however, tailor a given POVM to ones needs by suitable channel twirling to generate classical snapshots that capture relevant information of a given quantum state.  It would be interesting to investigate how well properties of quantum states can be learned for POVMs strongly different from the measurement of interest by combining our twirling technique with a classical learning strategy and explore where the boundaries of such schemes lie. 
We have restricted our attention to simple, idealized, models of CV POVMs and note that applying our methods to more accurate models of physical POVMs for tomography would be an interesting and potentially useful line of future work. Finally, we note that the projective nature of the twirling allows to adapt our base protocol to noisy measurement channels by following the strategy in ref.~\cite{Chen_2021}. In ref.~\cite{Chen_2021} the authors extract the effective logical channel parameters in a calibration experiment using known states prior to applying the shadow tomography protocol with suitably adapted postprocessing. This technique is readily applicable to our setup as well.

The probabilistic state decomposition derived from heterodyne measurements and displacement twirling relative to a GKP code also warrants further investigation. This protocol can be understood as a ``black-box'' scheme that produces relevant Gaussian samples from an arbitrary state in the input such that logical expectation values are aligned.  Since Gaussian states are computationally easy to tract through Gaussian evolutions,  this tool may find application in the simulation of realistically producible states with a priori unknown decomposition into Gaussian states as they are evolved under Gaussian evolutions which is relevant in the design of large scale experiments with the GKP code \cite{Bourassa_2021}. It would be interesting to investigate \textit{lower bounds} for this kind of decomposition: how many Gaussian states need to be mixed in order to emulate the logic of an encoded (GKP) state? This question, here motivated by our protocol and the proposed applications to simulation, motivate the general exploration of effective logical properties of easily accessible physical states which need not be restricted to the representation of encoded quantum states. It would further be interesting to understand what kind of logical dynamics can be effectively generated -- in expectation -- via Gaussian quantum channels.

Finally, we view the extension of the logical GKP shadow to a full-fledged CV shadow discussed in sec.~\ref{sec:chasing_shadows} as the most interesting contribution presented here. Here we proposed a random Wigner tomography scheme, where the statistical tools used in in shadow tomography~\cite{Toolbox} translate into a genuine continuous variable setting. We showed how logical expectation values of observables relative to a random GKP code can be used to reconstruct the full CV expectation value of said observable. In this part the application of tools from the theory of random lattices were the key element. 
It is interesting that random coding techniques are useful for (continuous variable) shadow tomography, which is an observation similarly made in ref.~\cite{conrad2025continuousvariabledesignsdesignbasedshadow}. 

A natural question is whether the techniques investigated here also apply to other quantum error correcting codes. At its core, the relevant ingredients to our base protocol are 1. a physical POVM that is complete on the code subspace, 2. a unitary realization of the logical Clifford group, 3. a measure over the set of representatives of the logical Clifford group. In the presented construction we used a measure that limits to a uniform distribution over \textit{all} representatives by including displacements over the full (infinite) stabilizer group. This choice ensured that the twirled POVM indeed results into a logical deoplarizing channel without logical coherent terms. While this aligns with the construction presented in ref.~\cite{Huang_2020}, this choice is not strictly necessary: any reduced set of Clifford representatives is expected to yield an invertible twirled POVM.

We hence expect our protocol to also apply to other bosonic codes such as the cat-~\cite{Legthas} or binomial codes~\cite{Michael}. We however do not expect the extrapolation to a physical shadow tomography protocol based on random codes to generalize as easily to other bosonic codes. For this protocol, our construction relies in particular on the mean value formula for symplectic lattices. It is conceivable that a similar effect can be obtained using quantum spherical codes~\cite{Jain_2024} by leveraging their intrinsic relation to spherical designs, which would be an interesting item of future research.
Another route would be to further investigate the extent to which the presented ideas are applicable to discrete variable systems, e.g., whether logical shadow tomography relative to random qubit-QECCs also allows to estimate properties of the full physical system. Here, it would be interesting to investigate the extractable information from elements of random stabilizer or subsystem codes on a multi-qubit quantum system through the lens of shadow tomography, where we speculate that techniques similar to those presented here may be applicable.
The toolbox developed here allows for a wide range of generalizations and possible applications,  for which we hope this work stimulates curiosity.
\medskip

\acknowledgments
We are grateful to %
V.~Albert, 
C.~Bertoni,
A.~Ciani,
J.~Haferkamp, 
J.~Iosue,
R.~Kueng,
J.~Magdalena de la Fuente,
Y.~Teng
and N.~Walk
for many inspiring and helpful discussions. 
JC thanks in particular C.~Bertoni for sharing his toolbox of ideas  on statistical bounds and V.~Albert and J.~Iosue for discussions on the relation between the techniques discussed in sec.~\ref{sec:chasing_shadows} and CV state designs. We further thank the anonymous referees for detailed and constructive feedback on this manuscript.
JC and JE gratefully acknowledge support from the BMFTR (RealistiQ, MUNIQC-Atoms, PhoQuant, QPIC-1, QSolid, PasQuops),  the DFG (CRC 183, SPP 2541, BoLaCo), the Munich Quantum Valley (K-8), 
the ERC (DebuQC), the Cluster of Excellence ML4Q, Quantum Berlin as well as the Einstein Research Unit on quantum devices.

\bibliographystyle{quantum}
\bibliography{Shadow_bib}

\begin{widetext}
    
\appendix

\section{Wigner function for general GKP states}\label{app:Wigner}
In this section we derive the Wigner function for a general idealized GKP state. Let $\rho$ be an arbitrary state and consider the projection of the state onto code space $\overline{\rho}=\Pi_{\CL}\rho \Pi_{\CL}$, which satisfies
\begin{equation}
    \forall \bs{\xi}\in \CL:\overline{\rho}=D\lr{\bs{\xi}}\overline{\rho}=\overline{\rho} D^{\dagger}\lr{\bs{\xi}},
\end{equation}
in particular this state is an operator in the commutant of the stabilizer group given by displacements in $\CL$ (see sec.~\ref{sec:twirl}). %
Let $\rho=\int_{\R^{2n}}d\bs{x}\,\rho\lr{\bs{x}} D\lr{\bs{x}}$. We have in the limit of a uniform distribution $P_{\CL}\lr{\bs{\xi}}\to $ 
\begin{equation}
    \overline{\rho}=\Pi_{\CL}\rho \Pi_{\CL} = \Pi_{\CL}\sum_{\bs{\xi}\in \CL} P_{\CL}\lr{\bs{\xi}} D\lr{\bs{\xi}} \rho D^{\dagger}\lr{\bs{\xi}},\label{eq:state_A2}
\end{equation}
where $P_{\CL}$ is a priori any probability measure supported on $\CL$ and the limiting constant is arbitrarily chosen. Using Poisson summation, we obtain
\begin{align}
\overline{\rho}&=|\det(\CL)|\Pi_{\CL}\sum_{\bs{\xi}^{\perp}\in \CL^{\perp}} \rho(\bs{\xi}^{\perp})  D(\bs{\xi}^{\perp})\\
&=|\det(\CL)|\Pi_{\CL}\sum_{[\bs{\xi}^{\perp}]\in \CL^{\perp}/\CL} \overline{\rho}([\bs{\xi}^{\perp}]) D([\bs{\xi}^{\perp}]) \label{eq:log_state}\\
&=|\det(\CL)|\sum_{\bs{\xi}^{\perp}\in \CL^{\perp}} \underbrace{\lrq{\sum_{\bs{\xi}\in \CL}\rho(\bs{\xi}^{\perp}-\bs{\xi})e^{-i\pi\bs{\xi}^TJ\bs{\xi}^{\perp}}}}_{c_{\bs{\xi}^{\perp}}}D(\bs{\xi}^{\perp}),
\end{align}
where the coefficients are such that $c_{\bs{\xi}^{\perp}+\bs{\xi}}=c_{\bs{\xi}^{\perp}}e^{-i\pi\bs{\xi}^TJ\bs{\xi}^{\perp}}$. In eq.~\eqref{eq:log_state} we have defined $[\cdot]:\CL \to \CL^{\perp}/\CL$, which reduces a point in $\CL$ to an element of a fixed choice of representatives of $\CL^{\perp}/\CL$ as well as the logical Bloch coefficients
\begin{equation}
    \overline{\rho}([\bs{\xi}^{\perp}]) =\sum_{\bs{\xi}\in \CL} \rho([\bs{\xi}^{\perp}]+\bs{\xi})e^{i\pi\bs{\xi}^TJ[\bs{\xi}^{\perp}]} = \overline{\rho}([\bs{\xi}^{\perp}]+\bs{\xi}')\; \forall \bs{\xi}'\in \CL.
\end{equation}

The Wigner function associated to this state is 
\begin{align}
    W_{\Pi_{\CL}\rho \Pi_{\CL}}\lr{\bs{x}}
    &= |\det(\CL)|\sum_{\bs{\xi}^{\perp}\in \CL^{\perp}} c_{\bs{\xi}^{\perp}} e^{-i2\pi \bs{x}^TJ\bs{\xi}^{\perp}}\nonumber\\
    &=|\det(\CL)|\sum_{[\bs{\xi}^{\perp}]\in \CL^{\perp}/\CL} \sum_{\bs{\xi} \in \CL} c_{[\bs{\xi}^{\perp}]+\bs{\xi}} e^{-i2\pi\bs{x}^TJ\lr{[\bs{\xi}^{\perp}]+\bs{\xi}}}\nonumber\\
    &=|\det(\CL)|\sum_{[\bs{\xi}^{\perp}]\in \CL^{\perp}/\CL}c_{[\bs{\xi}^{\perp}]} e^{-i2\pi\bs{x}^TJ[\bs{\xi}^{\perp}]} \sum_{\bs{\xi} \in \CL} e^{-i2\pi\lr{\bs{x}-[\bs{\xi}^{\perp}]/2}^TJ\bs{\xi}}\nonumber\\
    &= \sum_{[\bs{\xi}^{\perp}]\in \CL^{\perp}/\CL}c_{[\bs{\xi}^{\perp}]} e^{-i2\pi\bs{x}^TJ[\bs{\xi}^{\perp}]}\Sha_{\CL^{\perp}}\lr{\bs{x}-[\bs{\xi}^{\perp}]/2}.\label{eq:WignerGKP}
\end{align}
where again we have use Poisson summation, the Dirac comb $\Sha_{\CL^{\perp}}$ defined in the main text.
Eq.~\eqref{eq:WignerGKP} shows that, different from the characteristic function, the Wigner function can be supported on points outside of $\CL^{\perp}$. The only general restriction is that it no supports outside of $\CL^{\perp}/2$.

\section{Generating the symplectic group}

In this section, we discuss schemes for generating all elements in the symplectic 
group \cite{Continuous,Weedbrook_2012}.
Block matrices 
\begin{equation}
S=\begin{pmatrix}
A & B \\ C & D
\end{pmatrix} \in \Z_d^{2n \times 2n}
\end{equation}
are symplectic if $S^TJS=J$,  which requires $A^TC=C^TA$,  $B^TD=D^TB$ as well as$A^TD-C^TB=I$.  In particular, 
we have that for $B=C=0$ it is symplectic if $D=A^{-T}$ such that $S=A\oplus A^{-T}$.  If $A=D=I$  and $C=0$ ($B=0$) it becomes necessary that $B=B^T$ is symmetric ($C$ is symmectric). 
Matrices of these constrained types 
feature a particularly
simple structure and follow the simple multiplication rules
\begin{align}
\begin{pmatrix}
A_1 & 0 \\ 0 & A_1^{-T}
\end{pmatrix} \begin{pmatrix}
A_2 & 0 \\ 0 & A_2^{-T}
\end{pmatrix}
&=\begin{pmatrix}
A_1A_2 & 0 \\ 0 & \lr{A_1A_2}^{-T}
\end{pmatrix}, \\
\begin{pmatrix}
I & B_1 \\ 0 & I
\end{pmatrix} \begin{pmatrix}
I & B_2 \\ 0 & I
\end{pmatrix}
&=\begin{pmatrix}
I & B_1+B_2 \\ 0 & I
\end{pmatrix}, \\
\begin{pmatrix}
I & 0 \\ C_1 & I
\end{pmatrix} \begin{pmatrix}
I & 0 \\ C_2 & I
\end{pmatrix}
&=\begin{pmatrix}
I & 0 \\ C_1+C_2 & I
\end{pmatrix} .
\end{align}

In this section, we 
show specifically -- building on previous work on qubits \cite{AaronsonGottesman,  PatelMarkovHayes} -- how for prime dimension $q$,  symplectic matrices in $\Sp_{2n}\lr{\Z_d}$ can be synthesized from an elementary gate set $S=\lrc{ J_i ,   P_{i},  C_{i\rightarrow j}}$ of such constrained block matrices consisting of the following matrices in block form,   where $\pi_{i}=\bs{e}_i\bs{e}_i^T$ and $e_{i,j}=\bs{e}_i\bs{e}_j^T$:

\begin{itemize}
\item  The quantum Fourier transform on qudit $i$
\begin{equation} J_i=\begin{pmatrix}
I-\pi_{i} & \pi_i  \\ -\pi_i & I-\pi_i
\end{pmatrix},\, i\in \lrq{1,n}
\end{equation}
with $J_i^2=-I$,  mapping  $X_i \mapsto Z_i^{-1}, \, Z_i \mapsto X_i,$

\item the phase gate
\begin{equation}
P_i=\begin{pmatrix}
I & 0  \\ \pi_i& I
\end{pmatrix},\,  i\in \lrq{1,n},
\end{equation}
 mapping $X_i\mapsto X_i Z_i$ and 

\item the  CNOT gate
\begin{equation}
C_{i\rightarrow j}=\begin{pmatrix}
I+e_{j,i} &  0 \\ 0& I-e_{i,j}
\end{pmatrix},\,  i \neq j \in \lrq{1,n}
\end{equation}
 that maps $X_i \mapsto X_iX_j$.

\item The CNOT gate is of block diagonal form,  and it can be shown by performing the matrix multiplication that the upper triangular elementary block matrix
\begin{equation}
B_{i,j}=\begin{pmatrix}
I &  e_{i,j}+e_{i,j} \\ 0& I
\end{pmatrix} = J_j^{-1} C_{j\rightarrow i} J_j
\end{equation}
 mapping $Z_i\mapsto X_j Z_i$ and $Z_j\mapsto X_i Z_j$ can be obtained by conjugating the CNOT with a Hadamard type gate.  This generating set has $|S|=2n+n(n-1)$ elements,  where the contribution $n(n-1)$ comes from the fact that we assume all-to-all connectivity for the CNOTs in use.  This set can be reduced down to a set of $3n-1$ generators with CNOTs only between a linear number of pairs analogous to the Lickorish generators for the Dehn-twists mentioned in the main text,  which however would come at the cost of needing to mediate CNOTs not included in the set via a $O(n)$ number of those that are. 
 \end{itemize}

Denote sequences generated by a finite product from $S$ as $S^k:=\lrc{g_1 ,g_2,\hdots ,g_k,, \, g_i \,  \in S}$.
Similar to previous work on generating $\Sp_{2n}\lr{\Z_2}$ we show here a result for prime dimension.

\begin{lem}[Universality for prime dimensions]
Let $d$ be prime.  For the generating set $S={ J_i ,   P_{i},  C_{i\rightarrow j}}$ defined above,  we have
\begin{equation}
\Sp_{2n}\lr{\Z_d} \subseteq S^k
\end{equation}
for $k=O\lr{d n^2}$.
\end{lem}
That is,  sequences of $O(dn^2)$ of gates from $S$ suffice to generate all elements in $\Sp_{2n}\lr{ d}$.

\proof It has been shown in 
ref.~\cite{DopicoJohnson} that every symplectic matrix  $S\in \Sp_{2n}\lr{\Z_d}$,  $d$ prime admits a decomposition into symplectic matrices
\begin{equation}
S=Q\begin{pmatrix}
I & 0 \\ C & I
\end{pmatrix}
\begin{pmatrix}
A & 0 \\ 0& A^{-T}
\end{pmatrix}
\begin{pmatrix}
I & B \\ 0 & I
\end{pmatrix},
\end{equation}
where $A\in \mathrm{GL}\lr{n, d}$ is invertible and $C \in \mathrm{GL}_n\lr{d}$ and $B \in \mathrm{GL}_n\lr{d}$ are symmetric and $Q$ is a $O(n)$ length product of the matrices $J_i$ we have defined above. Reference \cite{DopicoJohnson}, in fact, has shown 
this for the field of complex numbers $\C$ but the proof carries over to any number field,  such as $\Z_d=\mathbb{F}_d$ for $d$ prime.  
Using this decomposition,  it suffices to check how each individual block matrix can be compiled from the generating set above.  Using that 
\begin{equation}
J\begin{pmatrix}
I & B \\ 0 & I
\end{pmatrix} J^T=\begin{pmatrix}
I & 0 \\ -B & I
\end{pmatrix}
\end{equation} together with $J=\prod_{i=1}^n J_i$ we have that the every upper block triangular matrix can be converted to a lower block triangular one with $O(n)$ overhead and that every block upper triangular matrix 
\begin{equation}
\begin{pmatrix}
I & B \\ 0 & I
\end{pmatrix}
\end{equation}
with $B=B^T$ can be obtained from a $O(dn^2)$ fold product of matrices of type $J_i P_i J_i^T$ and $B_{i,j}$ for their simple multiplication structure.  It remains to bound the complexity of compiling the block diagonal part $A\oplus A^{-T}$. Note that due to the simple multiplication structure of these matrices this problem is equivalent to bounding the complexity of compiling the blocks $A$ as generated by elements $I+e_{j,i}$.  This is bounded 
using 
the same argument 
as 
in 
ref.\  \cite{AaronsonGottesman}, which has employed 
a result from Patel 
et al.\  \cite{PatelMarkovHayes},  who have shown 
that for the underlying field $\Z_2$, an achievable lower bound is given by $O\lr{n^2/\log_2\lr{n}}$.  As was already noticed in 
ref.~\cite{PatelMarkovHayes},  their technique generalizes for any finite field with order $d$,  where it yields a bound $O\lr{n^2/\log_d\lr{n}}$.  In total,  we hence obtain a bound $O(dn^2)$ for the length of the product from $S$ to generate any element in $\Sp_{2n}\lr{\Z_d}$.
\endproof

\section{Proof of Theorem \ref{them:GKPshadow}}\label{app:GKPshadow_proof}

In this section, we provide a proof of Theorem~\ref{them:GKPshadow} 
of the main text.

\begingroup
\renewcommand\thethem{2}
\begin{them}[Full representation of the 
state]
Let $\CL \subset \R^{2n}$ denote the lattice corresponding to a scaled GKP code on $n$ modes that encodes $d^n$ logical dimension and let $\mu$ denote a measure over elements $\Aut^S\lr{\CL^{\perp}}$ forming a logical Clifford $2-$design and let $\CM=\SumInt dz \sketbra{\Pi\lr{z}}$ denote a physical POVM.   Let $\sket{\rho}$ be an arbitrary state on the $n$-mode Hilbert space. The state
\begin{equation}
\sket{S}=N^{-1} \sum_{i=1}^N \CU_{S,  i}\sket{\Pi\lr{z_i}}
\end{equation}
produced by sampling $N$ Gaussian unitary operations via the measure $\mu$ and measurement outcomes from the Born distribution $z_i \sim \sbraket{\Pi\lr{z}|\CU_{S,  i}^{\dagger}|\rho}$ approximates the logical value of the state in Hilbert-Schmidt distance

\begin{equation}
d_{\rm HS}\lr{\Dec \sket{\rho} ,  \Dec \sket{S}} \leq \delta_{\rm HS}^2
\end{equation}
with probability at least $1-\delta$ for 
\begin{equation}
N\geq \frac{2d^{2n}}{\alpha^2 \delta_{\rm HS}^2} \lrq{\ln\lr{\frac{2}{\delta}} + 2n \ln\lr{d}},
\end{equation}
where $\alpha$ is determined by the commutation of the twirled 
POVM with the decoder
\begin{align}
 \Dec\CM^{\tau} &=\alpha \Dec + \beta \sket{\Pi_{\CL} }.
\end{align}
In particular, 
we also have
\begin{equation}
\|\Dec \sket{\rho}-  \Dec \sket{S}\|_1 \leq d^{\frac{n}{2}}\frac{\delta_{\rm HS}}{2} . \label{eq:thm3_tr}
\end{equation}
\end{them}%
\endgroup

\proof
The random variable $x_i=\frac{1}{\alpha N}\sbraket{P_{\alpha} | \Dec  \CU_{S, i}|\Pi\lr{z_i}}$ lies in a range $\lrq{-1/\alpha N,\, 1/\alpha N}$, such that H{\"o}ffding's inequality implies that for all logical Pauli operators $P_{\alpha},\, \alpha=1,\hdots ,d^{2n}$ the probability 
of the arithmetic mean $\sbraket{P_{\alpha} | \Dec |S}$ to deviate from its expectation value is bounded by

\begin{equation}
    P\lr{|\sbraket{P_{\alpha} | \Dec |S} - \mathbb{E}\lrq{\sbraket{P_{\alpha} | \Dec |S}} | \geq \epsilon } \leq 2^{-N\alpha^2\epsilon^2/2},
\end{equation}
such that, requiring the rhs to be upper bounded by $\delta/d^{2n}$,
\begin{equation}
    N=\frac{2}{\alpha^2\epsilon^2}\lrq{\ln\lr{\frac{2}{\delta}}+2n\ln\lr{d}} \label{eq:B7}
\end{equation}
asserts that with probability $1-\delta$, the arithmetic mean recovers the true expectation value up to $\epsilon$ accuracy. In particular, this also bounds the logical Hilbert-Schmidt distance
\begin{equation}
    d_{\rm HS}\lr{\Dec \sket{\rho},\, \Dec \sket{S}} \leq d^{2n} \epsilon^2
\end{equation}
with probability $1-\delta$. Setting $\delta_{\rm HS}^2=d^{2n} \epsilon^2$ then yields the result. Finally, eq.\ \eqref{eq:thm3_tr} is recovered using the Cauchy-Schwarz inequality to trade off the trace distance with the Hilbert-Schmidt distance.
\endproof
The preceding proof features non-trivial sample complexity scaling in the number of modes $n$. The origin of this scaling is mainly the choice of norm we make as we aim at a full logical state-reconstruction. Firstly, in eq.~\eqref{eq:B7} we required that the error probability is bounded by $\delta/d^{2n}$. This guarantees an anti-concentration error bounded by $\delta$ for each of the $d^{2n}$ logical Pauli observables. The logarithmic scaling with the number of observables is consistent with the scaling obtained in ref.~\cite{Huang_2020}. Similarly, bounding the Pauli expectation value for each of the $d^{2n}$ logical Pauli operators by a value of $\epsilon$ implies only a $d^{2n} \epsilon$ bound on the Hilbert-Schmidt distance. Each of these contributions essentially arise from our ``abuse'' of the shadow tomography protocol to perform full tomography on the logical state, which requires well approximation of an exponentially large complete set of operators. Finally, we obtain a trade-off from Hilbert-Schmidt distance to trace distance.

\section{Proof of Lemma~\ref{lem:rdn_lat_points}}

In this section,
we prove Lemma~\ref{lem:rdn_lat_points}.

\begingroup
\renewcommand\thelem{1}
\begin{lem}[Random lattice point sampling]
    Let $\rho, G$ be operators such that the product of their Wigner functions $ \widetilde{G}\lr{\bs{x}}  =  W_{\rho}\lr{\bs{x}} G\lr{\bs{x}}$ is well defined, $\Tr\lrq{G^2}<\infty$ is finite and further assume that 
    $W_{\rho}\lr{\bs{x}}, G\lr{\bs{x}}$ 
    are Lipschitz-continuous, with $\|\nabla W_{\rho}\lr{\bs{x}}\| \leq l_{\rho}$ and $\|\nabla G\lr{\bs{x}}\| \leq l_{G}$. Set $\tilde{l}=l_{\rho}+l_G$.
    Let $\sigma \ll \lambda_1\lr{\CL^{\perp}}$ be a small parameter. It holds that
    \begin{align}
    \overline{G}^{\sigma}_{\CL^{\perp}} &=  \int d\bs{x}\, p_{\sigma}\lr{\bs{x};\CL^{\perp}}  \widetilde{G}\lr{\bs{x}}  \nonumber \\
    &=\sum_{\bs{\xi}^{\perp} \in \CL^{\perp}} \widetilde{G}\lr{\bs{\xi}^{\perp}} +\epsilon\lr{\sigma} \\
    &= \overline{G}_{\CL^{\perp}}+\epsilon\lr{\sigma} \nonumber
    \end{align}
    with 
    \begin{equation}
        |\epsilon\lr{\sigma}| \leq \sqrt{2}\sigma \frac{\sqrt{2} \Gamma\lr{n+\frac{1}{2}}}{  \Gamma\lr{n}} \tilde{l}.
    \end{equation}
    and furthermore
    \begin{align}
    \overline{G}^{\sigma, 2}_{\CL^{\perp}} &=  \int d\bs{x}\, p_{\sigma}\lr{\bs{x};\CL^{\perp}}  \widetilde{G}^2\lr{\bs{x}}  \nonumber \\
    &\leq \lr{\frac{2}{\pi\sigma^2}}^n\lr{\sigma}\Tr\lrq{G^2}.
    \end{align}
\end{lem}
\endgroup
   
\proof
We have
\begin{align}
    \overline{G}^{\sigma}_{\CL^{\perp}}=N_{\CL^{\perp}}^{-1} \sum_{\bs{\xi}^{\perp}\in \CL^{\perp}} e^{-\frac{\sigma^2}{2} \|\bs{\xi}^{\perp}\|^2} \int d\bs{x}\, e^{-\frac{1}{2\sigma^2}\|\bs{x}-\bs{\xi}^{\perp}\|^2} \widetilde{G}\lr{\bs{x}}.\label{eq:proof_dev_1}
\end{align}
In each summand, we have to first order
\begin{align}
    \int d\bs{x}\, e^{-\frac{1}{2\sigma^2}\|\bs{x}-\bs{\xi}^{\perp}\|^2} \widetilde{G}\lr{\bs{x}} 
   & = \int d\bs{x}\, e^{-\frac{1}{2\sigma^2}\|\bs{x}-\bs{\xi}^{\perp}\|^2} \widetilde{G}\lr{\bs{\xi}^{\perp}} + \int d\bs{x}\, e^{-\frac{1}{2\sigma^2}\|\bs{x}-\bs{\xi}^{\perp}\|^2} \lr{\bs{x}-\bs{\xi}^{\perp}}^T\nabla \widetilde{G}\lr{\bs{\xi}^{\perp}}.
\end{align}
The first term in this expression simply evaluates to 
\begin{equation}
    \int d\bs{x}\, e^{-\frac{1}{2\sigma^2}\|\bs{x}-\bs{\xi}^{\perp}\|^2} \widetilde{G}\lr{\bs{\xi}^{\perp}} = (2\pi \sigma^2)^n \widetilde{G}\lr{\bs{\xi}^{\perp}}
\end{equation}
and contributes the term $\overline{G}_{\CL^{\perp}}$ to the final expression, where the normalization factor $N_{\CL^{\perp}}$ perfectly cancels out.

Using Lipschitz continuity and 
the boundedness of the Wigner 
functions, it holds that $\|\nabla \widetilde{G}\lr{\bs{x}}\| \leq \tilde{l}$. Together with the Cauchy-Schwartz inequality, we can bound the second term
\begin{align}
\left|
\int d\bs{x}\, e^{-\frac{1}{2\sigma^2}\|\bs{x}-\bs{\xi}^{\perp}\|^2} \lr{\bs{x}-\bs{\xi}^{\perp}}^T\nabla \widetilde{G}\lr{\bs{\xi}^{\perp}} \right| &\leq \tilde{l} \int d\bs{x}\, e^{-\frac{1}{2\sigma^2}\|\bs{x}-\bs{\xi}^{\perp}\|^2} \left\|\bs{x}-\bs{\xi}^{\perp}\right\| \\
&=\sqrt{2}\sigma \lr{2\pi \sigma^2}^n\frac{\Gamma\lr{n+\frac{1}{2}}}{ \Gamma\lr{n}} \tilde{l}.
\nonumber
\end{align}
 Combining this with the normalization and sum yields an expression for the error
\begin{align}
    |\epsilon\lr{\sigma}| \leq \sqrt{2}\sigma \frac{\sqrt{2} \Gamma\lr{n+\frac{1}{2}}}{  \Gamma\lr{n}} \tilde{l}.
\end{align}
Finally, we use $|W_{\rho}\lr{\bs{x}}| \leq 2^n$ and $
    p_{\sigma}\lr{\bs{x};\CL^{\perp}} \leq \lr{2\pi\sigma^2}^{-n}$ to obtain
\begin{align}
\overline{G}^{\sigma, 2}_{\CL^{\perp}} &=  \int d\bs{x}\, p_{\sigma}\lr{\bs{x};\CL^{\perp}}  \widetilde{G}^2\lr{\bs{x}} \leq \lr{\frac{2}{\pi\sigma^2}}^n\int d\bs{x}\, G^2\lr{\bs{x}}=\lr{\frac{2}{\pi\sigma^2}}^n\Tr\lrq{G^2}.
\end{align}
\endproof

\section{Proof of Lemma~\ref{lem:scalar}}
Here, we prove the scalar product bound in Lemma~\ref{lem:scalar}. The main ingredient to the following derivation is a scalar product formula for functions on the Hilbert space associated to the space of lattices, $Y_n=\Sp_{2n}\lr{\Z}\backslash\Sp_{2n}\lr{\R}$, provided in ref.~\cite{Kelmer_2019}.

\begingroup
\renewcommand\thelem{2}
\begin{lem}[Scalar product bound]
    Let $f, g$ be two even compactly supported functions, it holds that

    \begin{equation}
        \left|\braket{\widetilde{F}_f, \widetilde{F}_g}_{Y_n}\right| \leq |\braket{f, 1}\braket{1, g}|+\frac{4\zeta\lr{n}^2}{\zeta\lr{2n}}\|f\|\|g\|.
    \end{equation}
\end{lem}
\endgroup

\proof
Define $f_k\lr{\bs{x}}=f\lr{k\bs{x}}$, $g_{k'}\lr{\bs{x}}=g\lr{k'\bs{x}}$. 
We have that
\begin{align}
    \left|\braket{\widetilde{F}_f, \widetilde{F}_g}_{Y_n} \right|&=\left|\sum_{k, k'\in \N} \braket{F_{f_k}, F_{g_{k'}}}\right| \\
    \nonumber
    &= \left|\sum_{k, k'\in \N}\frac{\braket{f_k, 1}\braket{1, g_{k'}}}{\zeta\lr{2n}^2} + \frac{2}{\zeta\lr{2n}}\lr{\braket{f_k, g_{k'}}+\braket{\iota \lr{f_k}, g_{k'}}}\right|\\
    \nonumber
    &=\left|\braket{f, 1}\braket{1, g} + \frac{2}{\zeta\lr{2n}}\sum_{k, k'\in \N}\lr{\braket{f_k, g_{k'}}+\braket{\iota \lr{f_k}, g_{k'}}}\right| \\
    \nonumber
    &\leq \braket{f, 1}\braket{1, g} +\frac{4}{\zeta\lr{2n}}\sum_{k, k'\in \N}\|f_k\| \|g_{k'}\|, 
    \nonumber
\end{align}
where we repeatedly use the triangle inequality, the fact that $\braket{f_k, 1}=k^{-2n} \braket{f, 1}, \braket{1, g_{k'}}=k'^{-2n} \braket{1, g} $ and the last line is derived using the Cauchy-Schwartz inequality $\braket{\iota \lr{f_k}, g_{k'}} \leq \|\iota \lr{f_k}\| \|g_{k'}\|=\|f_k\| \|g_{k'}\|$. 
Finally, note that it holds that
\begin{equation}
    \sum_k \|f_k\|=\sum_k \sqrt{\int d\bs{x} |f\lr{k\bs{x}}|^2} = \sum_k k^{-n}\sqrt{\int d\bs{x} |f\lr{\bs{x}}|^2} = \zeta\lr{n}\|f\|,
\end{equation}
which implies the final result. \endproof

\section{Proof of Lemma~\ref{lem:random_lat_1}}

\begingroup
\renewcommand\thelem{3}
\begin{lem}
    [Second moment: random lattice sampling]
    Let $\tilde{G}$ be an estimator as in Lemma~\ref{lem:rdn_lat_points} for an observable on a $n$-mode continuous variable quantum system,  symplectic  lattice $L=L^{\perp}\subset\R^{2n}$ and $d\in \mathbb{N}$ a natural number. Assume that the observable $G$ is such that 
    \begin{equation}
        \|G\|^2_2\coloneqq \int d\bs{x}\, |G\lr{\bs{x}}|^2<\infty 
    \end{equation}
    is finite. 
It holds that
\begin{equation}
    \int_{Y_n}d\mu\lr{L}\int d\bs{x}\, p_{\sigma}\lr{\bs{x}; d^{-1/2}L} \widetilde{G}\lr{\bs{x}}^2 \leq \lr{\frac{d}{2\pi}}^{n}\lrc{1+c_{\sigma, d}^n}\|G\|_2^2, 
\end{equation}
with $c_{\sigma, d}=\frac{2\pi d}{\sigma^2+\sigma^{-2}}\in [0, d\pi]$
\end{lem}
\endgroup

\proof
Note that from $|W_{\rho}\lr{\bs{x}}| \leq 2^n$ we generally have $\widetilde{G}\lr{\bs{x}}^2\leq 2^{2n}G\lr{\bs{x}}^2$. 
We can use the functional equation (Poisson summation) \cite{Conrad_2022} 
\begin{equation}
    \Theta_{\CL^{\perp}}\lr{z}=\det\lr{\CL}\lr{\frac{i}{z}}^n\Theta_{\CL}\lr{\frac{-1}{z}}
\end{equation}
to bound, using $\Theta_{\CL}\lr{i\frac{4\pi}{\sigma^2}}\geq 1$,
\begin{align}
N_{\CL^{\perp}}\lr{\sigma}=\det\lr{\CL^{\perp}}\lr{8\pi}^n\Theta_{\CL}\lr{i\frac{4\pi}{\sigma^2}} \geq \det\lr{\CL^{\perp}}\lr{8\pi}^n.
\end{align}
For lattices $\CL=\sqrt{d}L$, where $L$ is symplectic and $\det\lr{\CL}=d^n$ this yields the bound $N_{\CL^{\perp}}\lr{\sigma}^{-1}\leq \lr{d/8\pi}^{n}$. We thus estimate
\begin{align}
\int_{Y_n}d\mu\lr{L}\int d\bs{x}\, p_{\sigma}\lr{\bs{x}; d^{-1/2}L} \widetilde{G}\lr{\bs{x}}^2 &\leq \lr{\frac{d}{2\pi}}^{n}\int_{Y_n}d\mu\lr{L}\int d\bs{x}\, \lrc{e^{-\frac{1}{2\sigma^2}\|\bs{x}\|^2} + \widetilde{F}_{f_{\bs{x}}}} G\lr{\bs{x}}^2 \nonumber\\
&=\lr{d/2\pi}^{n}\int d\bs{x}\, \lrc{e^{-\frac{1}{2\sigma^2}\|\bs{x}\|^2} + \braket{\widetilde{F}_{f_{\bs{x}}}, 1}_{Y_n}}G\lr{\bs{x}}^2,
\end{align}
where we defined $f_{\bs{x}}\lr{\bs{v}}=e^{-\frac{\sigma^2}{2d}\|\bs{v}\|^2-\frac{1}{2\sigma^2}\|\bs{x}-d^{-1/2}\bs{v}\|^2}$.
Using the mean value formula, it holds that 
\begin{align}
    \braket{\widetilde{F}_{f_{\bs{x}}}, 1}_{Y_n}&=\int d\bs{v}\, e^{-\frac{\sigma^2}{2d}\|\bs{v}\|^2-\frac{1}{2\sigma^2}\|\bs{x}-d^{-1/2}\bs{v}\|^2}
    = \lr{\frac{2\pi d}{\sigma^2+\sigma^{-2}}}^n e^{-\frac{\Sigma}{2} \|\bs{x}\|^2}\eqqcolon c_{\sigma, d}^n e^{-\frac{\Sigma}{2} \|\bs{x}\|^2},
\end{align}
where we have called the expression in the final bracket $c_{\sigma, d}=\frac{2\pi d}{\sigma^2+\sigma^{-2}}$, which is bounded in the range $c_{\sigma, d}\in \lrq{0, d\pi}$ and have defined $\Sigma:=\sigma^{-2}\lr{1-\lr{1+\sigma^4}^{-1}}=\frac{\sigma^2}{1+\sigma^4} \approx \sigma^2$.
Note that $\forall s>0$, 
it holds that
\begin{align}
    \int d\bs{x}\,e^{-\frac{1}{2s^2}\|\bs{x}\|^2} G\lr{\bs{x}}^2 \leq \int d\bs{x}\,G\lr{\bs{x}}^2 \eqqcolon \|G\|_2^2.
\end{align}
Inserting this inequality yields the final bound
\begin{align}
    \int_{Y_n}d\mu\lr{L}\int d\bs{x}\, p_{\sigma}\lr{\bs{x}; d^{-1/2}L} \widetilde{G}\lr{\bs{x}}^2 \leq \lr{\frac{d}{2\pi}}^{n}\lrc{1+c_{\sigma, d}^n}\|G\|_2^2. 
\end{align}
\endproof

\section{Second moment bound for exact inner means}\label{app:lem4}
In this section, we prove the following lemma~\ref{lem:random_lat_inner}.
\begingroup
\renewcommand\thelem{4}
\begin{lem}[Second moment: random lattice sampling for exact inner means]
Let $\tilde{G}$ be an estimator as in Lemma~\ref{lem:rdn_lat_points} for an observable on a $n$-mode continuous variable quantum system with $n>1$, $L=L^{\perp} \subset \R^{2n}$ denote a symplectic lattice and let $d\in \mathbb{N}$ be a natural number.
 Assume that the observable $G$ is such that 
    \begin{equation}
        \|G\|_1 \coloneqq \int d\bs{x}\, |G\lr{\bs{x}}|<\infty 
    \end{equation}
    is finite. 
   Let  
   \begin{equation}
       \overline{G}_{\CL^{\perp}}^{\sigma}= \int d\bs{x}\, p_{\sigma}\lr{\bs{x};\, \CL^{\perp}} \widetilde{G}\lr{\bs{x}}
   \end{equation}
   be the estimator from Lemma~\ref{lem:rdn_lat_points} with $\CL^{\perp}=d^{-1/2}L$ and
    let $\sigma>0$. It holds that
   \begin{equation}
        \left|\int_{Y_n} d\mu\lr{L}\lr{\overline{G}_L^{\sigma}}^2\right|\leq  \lr{\frac{d}{4\pi}}^{2n}\lrq{1+\lr{2+2^{2-n}\frac{\zeta(n)^2}{\zeta(2n)}}c_{\sigma, d}^n+c_{\sigma, d}^{2n}}\|G\|_1^2
    \end{equation}
    with $c_{\sigma, d}=\frac{2\pi d}{\sigma^2+\sigma^{-2}}\in [0, d\pi]$
\end{lem}
\endgroup
\proof
We start by computing the second moment
    \begin{align}
    \int d\mu\lr{L}\lrq{\int d\bs{x}\, p_{\sigma}\lr{\bs{x};\, d^{-1/2}L} \widetilde{G}\lr{\bs{x}}}^2 &= \int d\mu\lr{L}\int d\bs{x}d\bs{y}\, p_{\sigma}\lr{\bs{x};\, d^{-1/2}L}p_{\sigma}\lr{\bs{y};\, d^{-1/2}L} \widetilde{G}\lr{\bs{x}}\widetilde{G}\lr{\bs{y}}.
    \end{align}
Swapping the order of integration, we first 
compute
\begin{align}
    \int d\mu\lr{L}p_{\sigma}\lr{\bs{x};\, d^{-1/2}L}p_{\sigma}\lr{\bs{y};\, d^{-1/2}L} &= \int d\mu\lr{L}N_{d^{-1/2}}\lr{\sigma}^{-2} \lrc{e^{-\frac{1}{2\sigma^2}\|\bs{x}\|^2}+\widetilde{F}_{f_{\bs{x}}} }\lrc{e^{-\frac{1}{2\sigma^2}\|\bs{y}\|^2}+\widetilde{F}_{f_{\bs{y}}} }, \nonumber \\
    &\leq \lr{\frac{d}{8\pi}}^{2n}\int d\mu\lr{L}\lrc{e^{-\frac{1}{2\sigma^2}\|\bs{x}\|^2}+\widetilde{F}_{f_{\bs{x}}} }\lrc{e^{-\frac{1}{2\sigma^2}\|\bs{y}\|^2}+\widetilde{F}_{f_{\bs{y}}}} \nonumber\\
    &= \lr{\frac{d}{8\pi}}^{2n}\left\{e^{-\frac{1}{2\sigma^2}\lr{\|\bs{x}\|^2+\|\bs{y}\|^2}}+ \braket{\widetilde{F}_{f_{\bs{x}}}, 1}_{Y_n} e^{-\frac{1}{2\sigma^2}\|\bs{y}\|^2}+\braket{1,\widetilde{F}_{f_{\bs{y}}}}_{Y_n} \right.
    \nonumber\\
    & \left. \times  e^{-\frac{1}{2\sigma^2}\|\bs{x}\|^2} + \braket{\widetilde{F}_{f_{\bs{x}}},\widetilde{F}_{f_{\bs{y}}}}_{Y_n}\right\}
\end{align}
where  we denote again $f_{\bs{x}}\lr{\bs{v}}=e^{-\frac{\sigma^2}{2d}\|\bs{v}\|^2-\frac{1}{2\sigma^2}\|\bs{x}-d^{-1/2}\bs{v}\|^2}$ and have used that 
$N_L\lr{\sigma}^{-1} \leq (d/8\pi)^{n}$ as in the previous proof. Here, we have
 inserted the relations from the proof of Lemma~\ref{lem:random_lat_1}.
Using Lemma~\ref{lem:scalar}, we also obtain the bound
\begin{align}
\left|\braket{\widetilde{F}_{f_{\bs{x}}},\widetilde{F}_{f_{\bs{y}}}}_{Y_n}\right| &\leq |\braket{f_{\bs{x}}, 1}\braket{1, f_{\bs{y}}}|+\frac{4\zeta\lr{n}^2}{\zeta\lr{2n}}\|f_{\bs{x}}\|\|f_{\bs{y}}\|.
\end{align}
We have already computed $\braket{f_{\bs{x}}, 1}=\braket{1, f_{\bs{y}}}$. Similarly, we obtain by completing the square
\begin{align}
    \|f_{\bs{x}}\|^2 = \int d\bs{v} e^{-\frac{\sigma^2}{d}\|\bs{v}\|^2 -\frac{1}{\sigma^2}\|\bs{x}-d^{-1/2}\bs{v}\|^2} = \lr{\frac{c_{\sigma, d}}{2}}^n e^{-\Sigma \|\bs{x}\|^2}.
\end{align}
 Taking everything together and using the triangle inequality gives rise to the bound
 \begin{align}
     \lr{\frac{8\pi}{d}}^{2n}\left|\int d\mu\lr{L}p_{\sigma}\lr{\bs{x};\, d^{-1/2}L}p_{\sigma}\lr{\bs{y};\, d^{-1/2}L}\right| 
     &\leq e^{-\frac{1}{2\sigma^2}\lr{\|\bs{x}\|^2+\|\bs{y}\|^2}}+ c_{\sigma, d}^n e^{-\frac{\Sigma}{2} \|\bs{x}\|^2} e^{-\frac{1}{2\sigma^2}\|\bs{y}\|^2}\nonumber\\
      &\hspace{1cm} +c_{\sigma, d}^n e^{-\frac{\Sigma}{2} \|\bs{y}\|^2}e^{-\frac{1}{2\sigma^2}\|\bs{x}\|^2}+ 
     c_{\sigma, d}^{2n} e^{-\frac{\Sigma}{2} \lr{\|\bs{x}\|^2+\|\bs{y}\|^2}} \nonumber\\
     &\hspace{1cm} +\frac{4\zeta\lr{n}^2}{\zeta\lr{2n}} 
     \lr{\frac{c_{\sigma, d}}{2}}^n e^{-\frac{\Sigma}{2}\lr{\|\bs{x}\|^2+\|\bs{y}\|^2}}.
 \end{align}
 
Again we use 
\begin{align}
    \int d\bs{x}\,e^{-\frac{1}{2s^2}\|\bs{x}\|^2}\tilde{G}\lr{\bs{x}} \leq   2^n\int d\bs{x}\,|G\lr{\bs{x}}|\eqqcolon 2^n\|G\|_1.
\end{align}
We can now combine all the above elements and obtain
\begin{align}
    \left|\int d\mu\lr{L}\lrq{\int d\bs{x}\, p_{\sigma}\lr{\bs{x};\,d^{-1/2}L} \widetilde{G}\lr{\bs{x}}}^2\right| &\leq 
    \lr{\frac{d}{4\pi}}^{2n}\lrq{1+\lr{2+2^{2-n}\frac{\zeta(n)^2}{\zeta(2n)}}c_{\sigma, d}^n+c_{\sigma, d}^{2n}}\|G\|_1^2.
\end{align}
\endproof

\section{Proof of Theorem~\ref{them:Randomlatshadow} and ~\ref{them:Randomlatshadow2}}
The following proofs are essentially restatements of the strategy followed in ref.~\cite{Huang_2020} with the appropriately adapted variance upper bounds.

\begingroup
\renewcommand\thethem{3}
\begin{them}[Random lattice CV shadows]
     Let $\tilde{\epsilon}, \delta, \sigma> 0$ be small parameters, let $G_{m},\, m=1,\hdots, M$ be operators with finite $\|G_m\|_2^2$ and set $K, B$ as described in protocol~\ref{protocol1}.

    $N=CKB$, samples from the distribution of phase space points $\lrc{\bs{x}_i}_{i=1}^N$ sampled according to protocol~\ref{protocol1} approximate the expectation values 
    \begin{equation}
        \overline{G}_{m}=\Tr\lrq{\rho G_m},
        m=1,\hdots,M
    \end{equation}
     of an arbitrary state on an $n$-mode continuous variable quantum system
     up to individual photon parity offsets $\tilde{G}_m\lr{0}$ and error $\tilde{\epsilon}$ 
     with probability at least $1-\delta$.
\end{them}
\endgroup
\proof

We proceed by a medians of means strategy equivalent that in ref.~\cite{Huang_2020}. 
Block the $N=KB$ estimates into $K$ batches each of size $B$. As per the result of Lemma~\ref{lem:rdn_lat_points} and using Chebychevs inequality the arithmetic mean $\widehat{G}_k$ of each of these batches approximates the final mean $\overline{G}$ up to an error of size $|\epsilon\lr{\sigma}|$ from Lemma~\ref{lem:rdn_lat_points} with failure probability
\begin{equation}
    P\lr{|\widehat{G}_k-\overline{G}| > |\epsilon\lr{\sigma}|+\tilde{\epsilon}} \leq \frac{V_1\lr{G; n,d,  \sigma}}{B\tilde{\epsilon}^2},
\end{equation}
with $V_1\lr{G; n, \sigma}$ as in lem.~\ref{lem:random_lat_1}.
Choosing $B\geq2^{2n}\lrc{1+\lr{\frac{2\pi}{\sigma^2}}^n} \|G\|_2^2 /\tilde{\epsilon}^2 $
and using Hoeffing's bound, the probability of deviation of the median of these estimates 
\begin{equation}
    G_{\rm MoM}\coloneqq {\rm median}\lrc{\widehat{G}_1,\hdots, \widehat{G}_K}
\end{equation}
from the real mean is bounded by
\begin{equation}
    P\lr{|G_{\rm MoM}-\overline{G}| \geq |\epsilon\lr{\sigma}|+\tilde{\epsilon}} \leq 2e^{-K/2}
\end{equation}
for all $\tilde{\epsilon}>0$. Choosing $K=2\log\lr{2M/\delta}$ suppresses the failure probability uniformly for $M$ observables.
\endproof

\begingroup
\renewcommand\thethem{4}
\begin{them}[Random lattice CV shadows with exact inner means]
     Let $\tilde{\epsilon}, \delta, \sigma> 0$ be small parameters, let $G_{m},\, m=1,\hdots, M$ be operators with finite $\|G_m\|_2^2$ and set $K, B$ as described in protocol~\ref{protocol2}.
    $N=CC'KBN_P$, samples from the distribution of phase space points $\lrc{\bs{x}_i}_{i=1}^N$ sampled according to protocol~\ref{protocol2} approximate the expectation values 
    \begin{equation}
        \overline{G}_{m}=\Tr\lrq{\rho G_m}+\tilde{G}_m\lr{0},
        m=1,\hdots,M
    \end{equation}
     of an arbitrary state on an $n$-mode continuous variable quantum system 
     up to a photon parity offset $\tilde{G}\lr{0}$ and error $\tilde{\epsilon}$ 
     with probability at least $1-\delta$.
\end{them}
\endgroup
\proof

We proceed by a medians of means strategy equivalent that in ref.~\cite{Huang_2020}. 
Block the $N=CKB$ estimates into $K$ batches each of size $B$. As per the result of Lemma~\ref{lem:rdn_lat_points} and using Chebychevs inequality the arithmetic mean $\widehat{G}_k$ of each of these batches approximates the final mean $\overline{G}$ up to an error of size $|\epsilon\lr{\sigma}|$ from Lemma~\ref{lem:rdn_lat_points} with failure probability
\begin{equation}
    P\lr{|\widehat{G}_k-\overline{G}| > |\epsilon\lr{\sigma}|+\tilde{\epsilon}} \leq \frac{V_1\lr{G; n,d,  \sigma}}{B\tilde{\epsilon}^2},
\end{equation}
with $V_1\lr{G; n, \sigma}$ as in lem.~\ref{lem:random_lat_1}.
Choosing $B\geq2^{2n}\lrc{1+\lr{\frac{2\pi}{\sigma^2}}^n} \|G\|_2^2 /\tilde{\epsilon}^2 $
and using Hoeffing's bound, the probability of deviation of the median of these estimates 
\begin{equation}
    G_{\rm MoM}\coloneqq {\rm median}\lrc{\widehat{G}_1,\hdots, \widehat{G}_K}
\end{equation}
from the real mean is bounded 
by
\begin{equation}
    P\lr{|G_{\rm MoM}-\overline{G}| \geq |\epsilon\lr{\sigma}|+\tilde{\epsilon}} \leq 2e^{-K/2}
\end{equation}
for all $\tilde{\epsilon}>0$. Choosing $K=2\log\lr{2M/\delta}$ suppresses the failure probability uniformly for $M$ observables.
\endproof

\end{widetext}

\end{document}